\documentclass[11pt]{article}
\usepackage{fullpage}
\usepackage{multirow}
\usepackage{cite}
\usepackage{graphicx}
\usepackage{amsmath}
\usepackage{amssymb}
\title{}
\date{}

\begin{document}
\bibliographystyle{utphys}
\def\beq{\begin{equation}}
\def\eeq{\end{equation}}
\def\e{\epsilon}
\def\bsp#1\esp{\begin{split}#1\end{split}}
\def\beqa{\begin{eqnarray}}
\def\eeqa{\end{eqnarray}}
\def\eqn#1{eq.~(\ref{#1})}
\def\bangle{\atopwithdelims \langle \rangle}

\newcommand{\msbar}{\ensuremath{\overline{\text{MS}}}}
\newcommand{\DIS}{\ensuremath{\text{DIS}}}
\newcommand{\abar}{\ensuremath{\bar{\alpha}_S}}
\newcommand{\bb}{\ensuremath{\bar{\beta}_0}}
\newcommand{\rc}{\ensuremath{r_{\text{cut}}}}
\newcommand{\Nd}{\ensuremath{N_{\text{d.o.f.}}}}
\setlength{\parindent}{0pt}

\titlepage
\begin{flushright}
Edinburgh 2013/27\\
IPPP/13/84\\
DCPT/13/168
\end{flushright}

\vspace*{0.5cm}

\begin{center}
{\Large \bf Webs and Posets}

\vspace*{1cm} \textsc{M. Dukes$^a$,
  E. Gardi$^b\footnote{Einan.Gardi@ed.ac.uk}$, H. McAslan$^c$,
  D. J. Scott$^d$ and
  C. D. White$^e$\footnote{Christopher.White@glasgow.ac.uk} } \\

\vspace*{0.5cm} $^a$ Department of Computer and Information Sciences, \\University of Strathclyde, Glasgow G1 1XH, UK\\

\vspace*{0.5cm} $^b$ Higgs Centre for Theoretical Physics, School of Physics and Astronomy, 
\\The University of Edinburgh, Edinburgh EH9 3JZ, Scotland, UK \\ 

\vspace*{0.5cm} $^c$ Department of Physics and Astronomy, University
of Sussex, Brighton BN1 9QH, UK\\

\vspace*{0.5cm} $^d$ Institute for Particle Physics Phenomenology, University of Durham, Durham DH1 3LE, UK\\

\vspace*{0.5cm} $^e$ SUPA, School of Physics and Astronomy,\\ University of Glasgow, Glasgow G12 8QQ, Scotland, UK\\

\end{center}

\vspace*{0.5cm}

\begin{abstract}
The non-Abelian exponentiation theorem has recently been generalised
to correlators of multiple Wilson line operators. The perturbative
expansions of these correlators exponentiate in terms of sets of
diagrams called webs, which together give rise to colour factors
corresponding to connected graphs.  The colour and kinematic degrees
of freedom of individual diagrams in a web are entangled by mixing
matrices of purely combinatorial origin. In this paper we relate the
combinatorial study of these matrices to properties of partially
ordered sets (posets), and hence obtain explicit solutions for certain
families of web-mixing matrix, at arbitrary order in perturbation
theory. We also provide a general expression for the rank of a general
class of mixing matrices, which governs the number of independent
colour factors arising from such webs. Finally, we use the poset
language to examine a previously conjectured sum rule for the columns
of web-mixing matrices which governs the cancellation of the leading
subdivergences between diagrams in the web. Our results, when combined
with parallel developments in the evaluation of kinematic integrals,
offer new insights into the all-order structure of infrared
singularities in non-Abelian gauge theories.
\end{abstract}

\vspace*{0.5cm}

\section{Introduction}
\label{sec:intro}
Wilson lines continue to generate significant amounts of interest, due
to their role in a variety of phenomenological and theoretical
applications~\cite{Arefeva:1980zd,Polyakov:1980ca,Dotsenko:1979wb,Brandt:1981kf,Korchemsky:1985xj,Ivanov:1985np,Korchemsky:1985xu,Korchemsky:1985ts,Korchemsky:1986fj,Korchemsky:1987wg,Korchemsky:1988hd,Korchemsky:1988si,Collins:1989bt,Korchemsky:1991zp,Kidonakis:1998nf,Kidonakis:1997gm,Kidonakis:1996aq,Kidonakis:2010dk,Drummond:2007cf,Basso:2007wd,Alday:2007hr,Pestun:2007rz,Drukker:2012de,Chien:2011wz,Cherednikov:2012qq,Cherednikov:2012yd,Henn:2013wfa,Naculich:2011ry,White:2011yy,Akhoury:2011kq,Miller:2012an,Beneke:2012xa}. For
example, they govern the structure of infrared singularities in
scattering amplitudes, which lead to large kinematic logarithms which
have to be summed up to all orders in perturbation
theory~\cite{Collins:1989gx,Korchemsky:1992xv,Korchemsky:1993uz,Catani:1996yz,Kidonakis:1998nf,Kidonakis:1997gm,Kidonakis:1996aq,Oderda:1999kr,Bonciani:1998vc,Beneke:2009rj,Beneke:2009ye,Ahrens:2010zv,Czakon:2013goa,Bauer:2000ew,Bauer:2000yr,Bauer:2001ct,Bauer:2001yt,Bauer:2002nz,Becher:2006nr,Becher:2006mr,Becher:2007ty,Yennie:1961ad,Sterman:1986aj,Catani:1989ne,Laenen:2008gt}.
They also govern scattering in the high energy limit in both gauge
theories and
gravity~\cite{Sotiropoulos:1993rd,Korchemsky:1993hr,Korchemskaya:1994qp,Korchemskaya:1996je,Balitsky:1995ub,Kovchegov:1996ty,Balitsky:2001gj,Balitsky:2009yp,JalilianMarian:1996xn,Gardi:2006rp,DelDuca:2011xm,DelDuca:2011ae,Kovchegov:2012mbw,Mueller:1993rr,Melville:2013qca,Akhoury:2013yua,Ware:2013zja}. For
these and other applications, it is important to classify the
behaviour of correlators of products of Wilson line operators. Whilst
much is known about the case of two Wilson lines meeting at a
cusp~\cite{Gatheral:1983cz,Frenkel:1984pz,Sterman:1981jc}, the general
case of more than two Wilson lines has been studied only recently. The
anomalous dimension has been calculated up to two-loop order for both
massless~\cite{Aybat:2006wq,Aybat:2006mz} and
massive~\cite{Ferroglia:2009ep,Ferroglia:2009ii,Mitov:2010xw}
particles, where in the former case it was found to have the same
colour structure as the one-loop result. This was later explained by
constraints on the structure of infrared singularities in massless
scattering amplitudes which follow from factorisation and rescalaing
symmetry, leading to the formulation of the dipole
formula~\cite{Becher:2009cu,Becher:2009qa,Gardi:2009qi}), a minimal
all-order ansatz for the anomalous dimension. Possible corrections to
this form have been investigated further
in refs.~\cite{Dixon:2009ur,DelDuca:2011xm,DelDuca:2011ae,Ahrens:2012qz,Naculich:2013xa,Caron-Huot:2013fea}. Further
progress in this area requires explicit calculations of Wilson line
correlators at higher loop orders. In practice this means calculating
the exponents of correlators of Wilson lines, from which the relevant
anomalous dimensions can be extracted. This itself necessitates the
development of efficient computational techniques.\\

In a recent series of
papers~\cite{Gardi:2010rn,Gardi:2011wa,Gardi:2011yz,Gardi:2013ita}
(see also~\cite{Mitov:2010rp}), a diagrammatic method has been
developed, which allows us to calculate the exponents of Wilson line
correlators directly. This provides an efficient framework in which to
address higher loop contributions to the multileg soft anomalous
dimension, and is a generalisation of the well-known diagrammatic
exponentiation of correlators of two Wilson lines (or a Wilson loop)
in terms of {\it
  webs}~\cite{Gatheral:1983cz,Frenkel:1984pz,Sterman:1981jc}. As is
explained in detail in ref.~\cite{Gardi:2010rn}, however, this
generalisation is highly non-trivial. Whereas webs in two-parton
scattering are single irreducible diagrams, which are each
individually free of subdivergences associated with renormalisation of
the cusp, in multi-parton scattering webs are closed sets of diagrams
related by permutations of gluon emissions on the Wilson lines (we
review this material in more detail in section~\ref{sec:review}). The
general form of a given correlator of Wilson line operators $\Phi_i$
is then given by
\begin{equation}
\left\langle\prod_{i=1}^L\Phi_i\right\rangle=\exp\left[\sum_W W\right],
\label{Wilsonprod}
\end{equation}
where the contribution of each web is 
\begin{equation}
W=\sum_{D,D'}{\cal F}(D)R_{DD'}C(D').
\label{Wform}
\end{equation}
Here the sums are over all diagrams $D$, $D'$ in the web, whose kinematic
and colour parts are given by ${\cal F}(D)$ and $C(D')$
respectively. The matrix $R_{DD'}$ consists of rational numbers of
combinatorial origin, and acts to entangle the colour and kinematic
degrees of freedom. Each web has its own web-mixing matrix, and
these are known to have special properties. Chief among these is
idempotence (proven in ref.~\cite{Gardi:2011wa}), which implies that web
mixing matrices are projection operators, with eigenvalues in the set
$\{0,1\}$. Each unit eigenvalue is associated with a combination of
kinematic factors which survives in the exponent, and which has an
associated {\it exponentiated colour factor} (ECF), consisting of a
superposition of colour factors of individual diagrams in the web. It
has recently been shown that all such ECFs correspond to colour
factors of diagrams in which all gluons are connected (once the Wilson
lines have been removed), thus generalising the so-called non-Abelian
exponentiation theorem from the two-line to the multiline
case~\cite{Gardi:2013ita}.\\

Other properties of web-mixing matrices are known, such as the fact
that their rows sum to zero~\cite{Gardi:2010rn,Gardi:2011wa}, and that
their columns appear to sum to zero after contraction with appropriate
weight factors~\cite{Gardi:2011yz}. These features are related to two
underlying principles: (a) colour factors in the exponent must be
fully connected (as described above); (b) the corresponding
combinations of kinematic factors are only allowed to contain specific
higher-order divergences, which can be cancelled by highly constrained
subtraction terms involving subdiagrams of the given web
(see~\cite{Gardi:2011yz}). It is therefore clear that a deeper
understanding of the combinatorics of web-mixing matrices can provide
a valulable insight into the structure of infrared singularities in
non-Abelian gauge theories, to all orders in perturbation theory. This
is only part of the story, however. A full understanding of higher
order singularity structures requires calculation of the necessary
kinematic integrals. To this end, new computational techniques are
becoming available, which offer the possibility of calculating certain
families of kinematic integral to any order in perturbation
theory~\cite{integrals,integrals2}. Clearly, analysis of the structure of web
mixing matrices and that of kinematic integrals should proceed in
parallel.\\

The aim of this paper is to investigate the combinatorial structure of web
mixing matrices in more detail. We will present a useful
classification of webs based on the number of external lines they
connect before focusing on the special case of webs which connect
$L+1$ lines with $L$ single gluon exchanges. We will present a simple
formula for the rank of any web-mixing matrix in this class and, thus,
for the number of independent connected colour factors such a web
contributes to. We will then examine the results
of ref.~\cite{Dukes:2013wa}, a recent study which considered web-mixing
matrices from a pure mathematical point of view, and which related
their combinatorics to that of order-preserving maps on partially
ordered sets (posets), for which a large mathematical literature
exists. Using the poset language, we will be able to completely solve
for the web-mixing matrices for two special families of webs, at
arbitrary order in perturbation theory. The calculation of the
corresponding kinematic integrals will be considered
elsewhere~\cite{integrals,integrals2}. Finally, we will interpret a recently
conjectured column sum rule for web-mixing
matrices~\cite{Gardi:2011yz} in the poset language, and look at this
in detail for the two particular web families whose web-mixing
matrices we are able to exactly solve for.\\

Although much about web-mixing matrices and their associated kinematic
combinations remains unexplored (particularly for webs which contain
three and four-gluon vertices away from the Wilson lines), their
classification through combinatorial methods offers new insights into
the structure of non-Abelian gauge theory. The web-mixing matrix
solutions we present here correspond to all-order properties in the
{\it exponents} of scattering amplitudes, thus correspond to very deep
structures in perturbation theory. There is a clear motivation for
pushing such investigations to broader web families, in tandem with
the calculation of the relevant kinematic integrals. \\

The structure of the paper is as follows. In section~\ref{sec:review}
we review the definition of web-mixing matrices, in a self-contained
fashion. In particular, we give a combinatorial definition of web
mixing matrix elements (first presented in ref.~\cite{Gardi:2011wa}) that
will be used frequently throughout the rest of the paper. In
section~\ref{sec:rank} we consider how to systematically classify
webs, and provide an expression for the rank of a particular class of
webs, as mentioned above. In section~\ref{sec:posets}, we introduce
terms and properties relating to posets, which will not necessarily be
familiar to a particle physics audience. In section~\ref{sec:results}
we consider two particular classes of webs, and use the poset language
to obtain explicit solutions for their associated web-mixing matrices,
before verifying that they encode fully connected colour factors. In
section~\ref{sec:colsum} we discuss the interpretation of the weighted
column sum rule of ref.~\cite{Gardi:2011wa} in terms of linear extensions,
and verify that this holds for the two web families considered in
section~\ref{sec:results}. In section~\ref{sec:conclude} we discuss
our results and conclude.

\section{Webs and web-mixing matrices}
\label{sec:review}

The aim of this section is to briefly review previous results on webs
and their mixing matrices, which will be useful for the rest of the
paper. Note that different conventions exist in the literature for
what is meant by a multiparton web~\cite{Gardi:2010rn,Mitov:2010rp}
(see in particular appendix A of ref.~\cite{Gardi:2011wa} for a discussion
of this point). Throughout this paper, we adopt the terminology of
refs.~\cite{Gardi:2010rn,Gardi:2011wa, Gardi:2011yz,Gardi:2013ita}. \\

\begin{figure}
\begin{center}
\scalebox{0.7}{\includegraphics{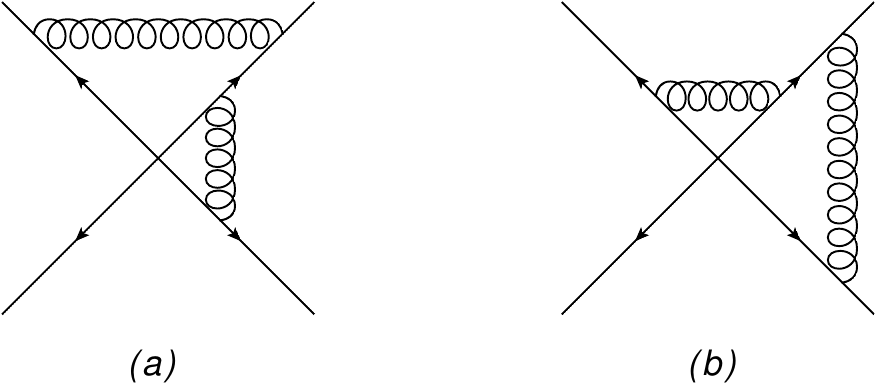}}
\caption{A (1,2,1) web, connecting three parton lines at two-loop order.}
\label{fig:2loop}
\end{center}
\end{figure}

A {\it multiparton web} is a set of diagrams contributing to the
exponent of a vacuum expectation value of Wilson line operators
meeting at a point (alternatively, to the soft gluon part of a QCD
scattering amplitude). An example is shown in figure~\ref{fig:2loop},
which illustrates a web connecting three Wilson lines (out of a total
of four) at two-loop order. This consists of a pair of diagrams, which
are related by permutations of gluon attachments on the external
lines.  Furthermore, the set is closed under such permutations:
permuting the gluons in diagram (a) leads to diagram (b), and vice
versa. In general, a web is any such set of diagrams closed under
permutations of gluon attachments.  Note that the two-loop web of
figure~\ref{fig:2loop} consists of distinct gluon exchanges. However,
there could also be three and four-gluon vertices off the eikonal
lines. This does not affect our definition of a web, which only
concerns permutations of gluon attachments associated with the hard
external (Wilson) lines. Many examples of webs can be found in
refs.~\cite{Gardi:2010rn,Gardi:2011wa,Gardi:2011yz,Gardi:2013ita}, and
we will see more throughout the present paper. Note that any diagram
that consists of a single connected piece (when the Wilson lines are
removed) is a web by itself; we will not be interested in such webs in
this paper as they do not present any combinatorial problem.\\

We may label a given web using a notation $(n_1,n_2\ldots n_L)$, where
$n_i\geq 0$ is the number of gluon attachments on parton line $i$, and
it is assumed that there are $L$ lines in total. Note that this
notation does not uniquely specify a given web. In particular, it does
not specify possible attachments of gluons off the eikonal lines, via
three and four-gluon vertices. This will nevertheless be a convenient
notation in what follows, since our interest here is in multiple gluon
exchange diagrams with no three or four gluon vertices. \\

\begin{figure}
\begin{center}
\scalebox{0.8}{\includegraphics{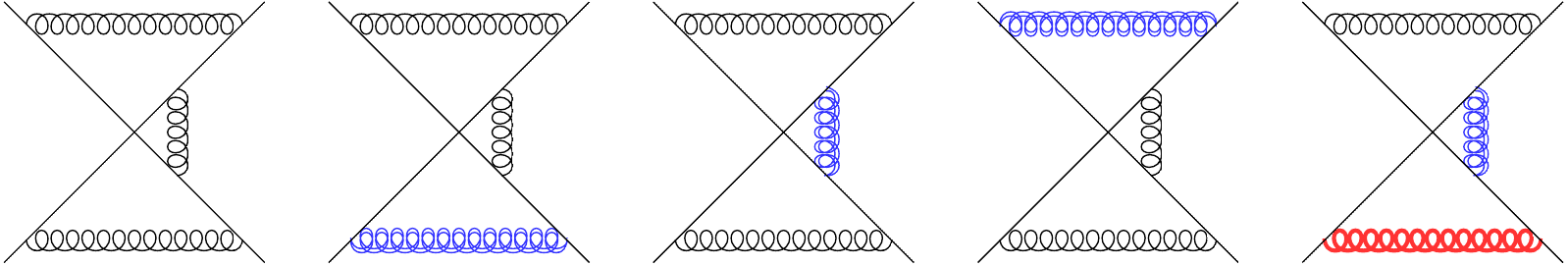}}
\caption{The partitions of the diagram on the far left-hand side, with
  $m=1,2,2,2,3$ respectively. Note that we have used different symbols
  to distinguish gluons belonging to different partition elements
  (double line, thick line), in addition to colours.}
\label{partex}
\end{center}
\end{figure}

\begin{figure}
\begin{center}
\scalebox{0.4}{\includegraphics{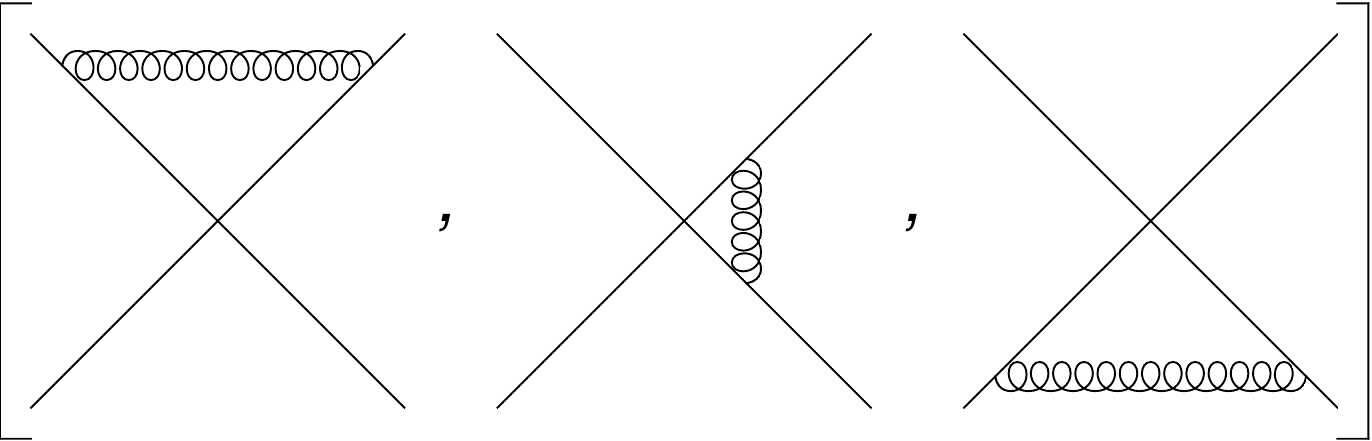}}
\caption{Alternative notation for the 3-partition of figure~\ref{partex},
in which each element of the partition is represented by its own diagram.}
\label{partex2}
\end{center}
\end{figure}

Given a web $W$, we may consider all diagrams $D\in W$. Each diagram
$D$ has a kinematic part ${\cal F}(D)$ and a colour factor
$C(D)$. Then the contribution of $W$ to the exponent of the Wilson
line correlator is given by eq.~(\ref{Wform}), where $R_{DD'}$ is a
{\it web-mixing matrix}. The existence of web-mixing matrices was
first derived using statistical physics methods (the ``replica
trick''), which also provides an algorithm for how to calculate
them~\cite{Gardi:2010rn}. One may also provide a closed-form
combinatorial formula for the element $R_{DD'}$. To this end, one
needs the notion of a {\it partition} of a web diagram $D\in W$.  An
{\it $m$-partition of diagram $D$} is a colouring of $D$ with a number
of distinct colours $m\leq n_c$, where $n_c$ is the number of
connected subdiagrams~\footnote{By a {\it connected subdiagram}, we
  mean a part of the diagram that forms a single connected piece, if
  the external (Wilson) lines have been removed.} in $D$. This is
illustrated in figure~\ref{partex}, which shows an example web diagram
at three-loop order together with its partitions (note that the
diagram itself is an $m$-partition with $m=1$).
Although thinking of partitions as colourings gives an easy way to
represent them, we actually use another notation to represent
partitions: an $m$-partition of a diagram $D$ will be represented as a
set with $m$ elements, each of which consists of the usual diagram
notation, but only including gluons of a particular colour. To
illustrate this, consider the 3-partition on the far right of
figure~\ref{partex}. We may represent this using the alternative
notation of figure~\ref{partex2}. Here, each separate colour in the
3-partition gets its own diagram, and there are thus 3 elements in the
set. Furthermore, we can then dispense with the colours altogether, as
these are no longer needed to separate out the elements of the
partition.\\

\begin{figure}
\begin{center}
\scalebox{0.6}{\includegraphics{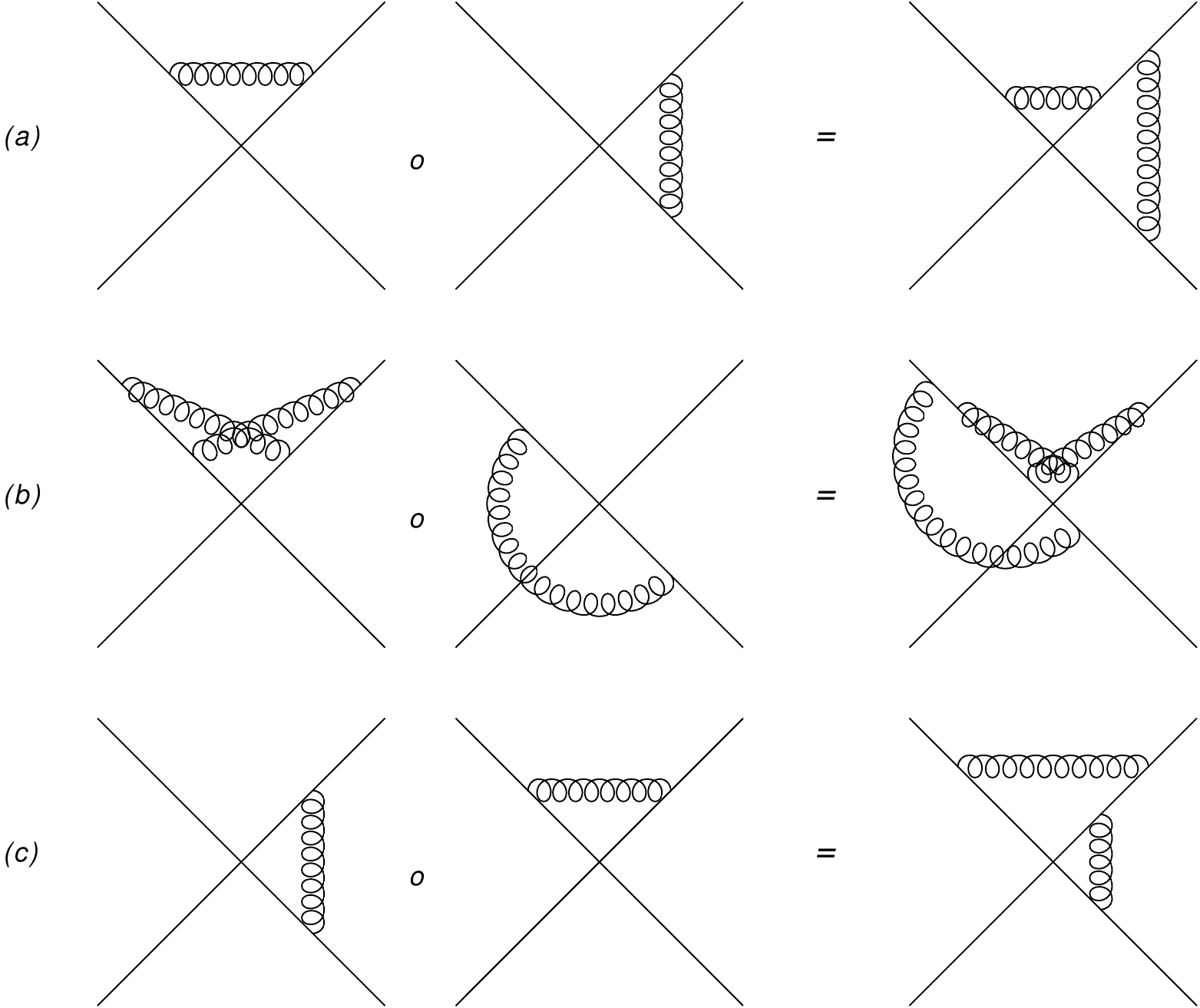}}
\caption{Examples of the multiplication rule $D_1\circ D_2$ for
  diagrams.}
\label{multex}
\end{center}
\end{figure}

The reason for using this alternative notation for partitions is that the 
diagrams are subject to a multiplication rule, which we have illustrated
in figure~\ref{multex}. 
The rule for combining diagrams is as follows: to form the product,
one takes the gluon part of each diagram, and draws them successively
emanating outwards from the center of the product diagram. In
figure~\ref{multex}(a), for example, the gluon joining external lines
1 and 2 (counting clockwise from top-left) appears closer to the
center than the gluon joining external lines 2 and 3 in the product.
Another example is shown in figure~\ref{multex}(b). Here the first
diagram on the left-hand side itself has two gluons. In the product
diagram, this entire structure is reproduced, followed by the gluon
which joins lines 1 and 3. One may also think of this rule as
operating separately on the external lines: on each external line, one
must order the gluon attachments outwards from the center of the
diagram, according to the order in which they appear in the product of
individual diagrams on the left-hand side. The reason for considering
this multiplication rule is that it describes how the colour factor of
a composite diagram is formed from the colour factors of its
constitutents. That is, denoting the product of diagrams $D_1$ and
$D_2$ via $D_1\circ D_2$, the colour factor of the product is given by
\begin{equation}
C(D_1\circ D_2)=C(D_1)C(D_2).
\label{Cprod}
\end{equation}
Note that this multiplication rule for diagrams is non-commutative,
corresponding to the fact that colour matrices associated with gluon
emissions are themselves non-commuting. For example, the alternative
ordering of the diagrams in figure~\ref{multex}(a) gives the product
shown in figure~\ref{multex}(c), which is indeed not the same. Rather,
it is a different diagram belonging to the same web.\\

\begin{figure}[h]
\begin{center}
\scalebox{0.8}{\includegraphics{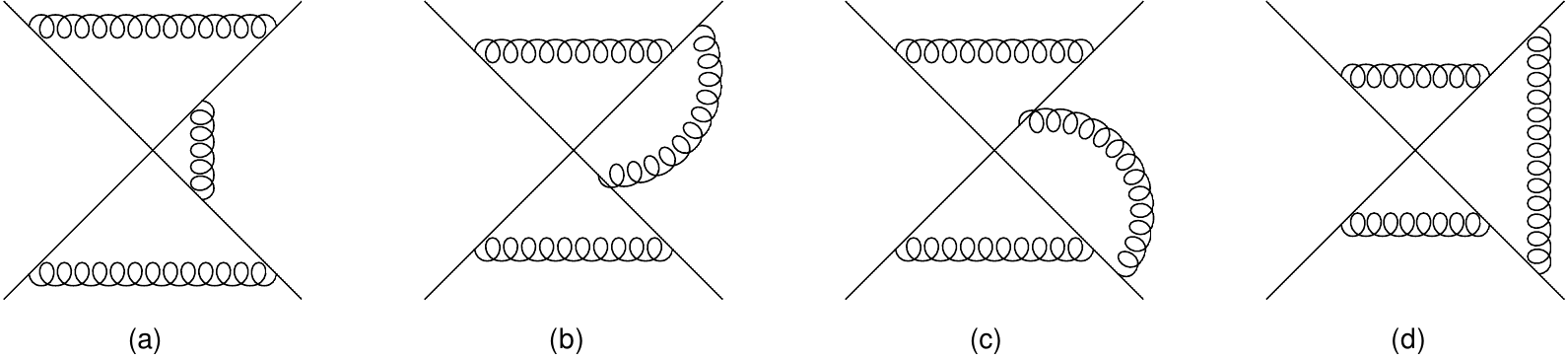}}
\caption{Three-loop web containing the diagram of figure~\ref{partex}.}
\label{webex}
\end{center}
\end{figure}

\begin{figure}
\begin{center}
\scalebox{0.6}{\includegraphics{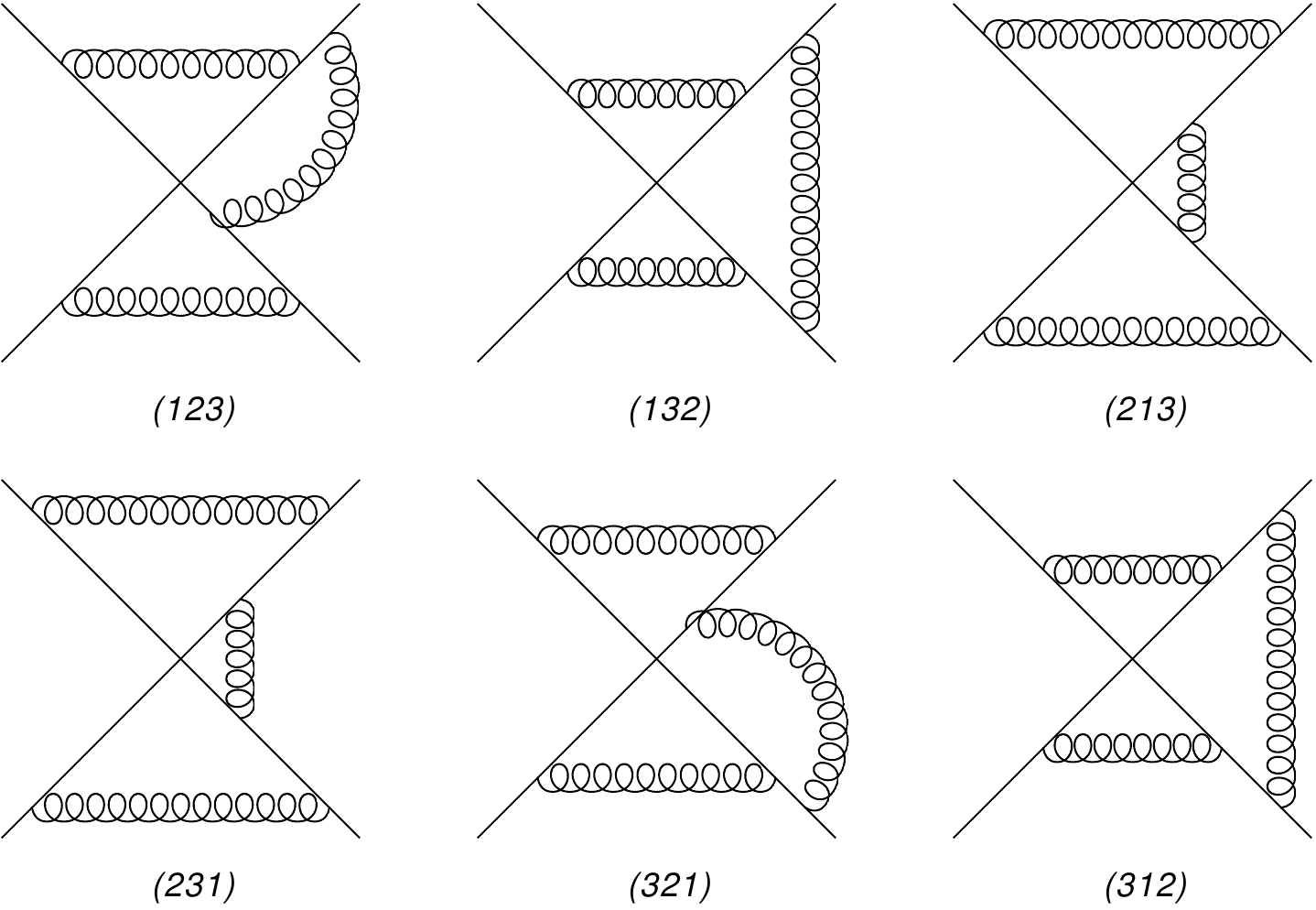}}
\caption{Diagrams formed by multiplying all permutations of elements in the 
3-partition of figure~\ref{partex2}.}
\label{multpart}
\end{center}
\end{figure}

Having defined the above multiplication rule, we can apply it within 
$m$-partitions of a given diagram, by multiplying the elements together. A 
given $m$-partition can be used to make $m!$ diagrams (where all elements of 
the partition are used in each product), according to the number of 
permutations of the elements. Note that some of the diagrams one forms may be 
the same. Consider, for example, the 3-partition from figure~\ref{partex2}.
This can be used to make 6 diagrams, which must belong to the web shown in
figure~\ref{webex}. Labelling the elements of 
figure~\ref{partex2} by [1,2,3], the diagrams that result from each 
permutation are shown in figure~\ref{multpart}.
Comparing this with figure~\ref{webex}, we see that diagrams (a) and (d) are 
both made twice, whereas diagrams (b) and (c) are made once only. The sum of 
these values is 2+2+1+1=3! as required. One may carry out a similar exercise 
for any $m$-partition of any diagram, and this leads us to define the 
following quantity, first introduced in ref.~\cite{Gardi:2011wa}:\\

{\it The ``overlap function of $D'$ with respect to $P_D$'',
 denoted by $\langle D'|P_D\rangle$,
is the number of ways that diagram $D'$ is made upon multiplying together all permutations
of the elements of the partition $P_D$ of diagram $D$. Here $D$ and $D'$ are diagrams in the
same closed set (related by gluon permutations).}\\

In the example of figure~\ref{multpart}, one finds 
$\langle d|P\rangle=\langle a|P\rangle=2$ and 
$\langle b|P\rangle=\langle c|P\rangle=1$, where $P$ is the 3-partition of 
diagram (a), shown in figure~\ref{partex2}. Note that, in general, one has
\begin{equation}
\sum_{D'}\langle D'|P_D\rangle =n(P_D)!,
\label{sumoverlap}
\end{equation}
for any fixed partition $P_D$, where $n(P_D)$ is the number of elements in 
this partition. That is, every permutation of the elements of $P_D$ makes 
some diagram in the set to which $D$ and $D'$ belong. Thus, summing over the 
overlap functions, by their definition, leads to the total number of 
permutations of the elements, which is $n(P_D)!$. \\

Given the notion of an overlap function, one may show that the web 
mixing-matrix element $R_{DD'}$ is given by~\cite{Gardi:2011wa}
\begin{equation}
R_{DD'}=\sum_{P_D}\frac{(-1)^{n(P_D)-1}}{n(P_D)}\langle D'|P_D\rangle.
\label{Rdef}
\end{equation}
There is a sum over all partitions $P_D$ of a given diagram $D$ on the
right-hand side, such that the left-hand side still depends on $D$
(and indeed on $D'$ which appears in the overlap function in each term
on the right-hand side). As already discussed in the introduction,
this formula is purely combinatorial in nature, and forms a basis for
the formal study of web-mixing matrices~\cite{Dukes:2013wa}. \\

Some general properties of web-mixing matrices have already been established.
Firstly, all web-mixing matrices are idempotent:
\begin{equation}
\sum_{E}R_{DE}R_{ED'}=R_{DD'}.
\label{idemp}
\end{equation}
This means that web-mixing matrices are projection operators, whose
eigenvalues are 0 and 1 (with associated multiplicities). This fact
has been proven using statistical physics methods~\cite{Gardi:2011wa},
but a combinatorial proof remains elusive. The physical interpretation
of the idempotence property was examined in ref.~\cite{Gardi:2010rn} and
then in detail in refs.~\cite{Gardi:2011yz,Gardi:2013ita}. Unit eigenvalues
of a web-mixing matrix are associated with certain combinations of
kinematic factors, each of which has an accompanying colour factor
which corresponds to that of a fully connected gluon
diagram~\cite{Gardi:2013ita}. Note in particular that the idempotence
property implies that the rank of any web-mixing matrix is the same as
its trace. The physical meaning of the rank is that it specifies the
number of independent connected colour factors to which a given web
contributes.\\

A second known property is that the row sums of web-mixing matrices
are always zero:
\begin{equation}
\sum_{D'} R_{DD'}=0\quad\forall\, D.
\label{zerosum}
\end{equation}
This was proven using combinatorial methods in ref.~\cite{Gardi:2011wa}, and this
property expresses the fact that the fully symmetric colour factor that can
be obtained from a given web does not contribute to the exponent of the 
amplitude. Rather, it is generated in the amplitude itself by the 
exponentiation of lower-order contributions. Also in ref.~\cite{Gardi:2011wa}, the
sum rule was generalised for planar and non-planar subsets of diagrams in a 
given web. This provides an interesting link between the combinatorial 
properties of web-mixing matrices, and the $1/N_c$ expansion, where $N_c$ is
the number of colours. \\

Thirdly, all web-mixing matrices are conjectured to obey the weighted column
sum rule:
\begin{equation}
\sum_{D}s(D)R_{DD'}=0\quad\forall\, D',
\label{column}
\end{equation}
where $s(D)$ is the number of ways of independently shrinking
connected subdiagrams to the origin, and such that $s(D)=0$ for
diagrams that are not maximally reducible.  For example, in the
diagrams of figure~\ref{webex}(b) and (c), there is only one way in
each graph of shrinking all three gluons independently to the origin,
whereas in (a) and (d), there are two ways, depending on whether one
shrinks the upper or lower gluon first. Thus, in this case the vector
of multiplicity factors is given by (2,1,1,2). More examples of the
$s(D)$ factors were presented in ref.~\cite{Gardi:2011wa}, and a
justification of the column conjecture given for graphs which do not
contain three or four gluon vertices. The physics of this result
concerns the renormalisation of the hard interaction vertex at which
the Wilson lines meet, and the physical justification of
eq.~(\ref{column}) relied upon the fact that individual webs
renormalise independently, with no mixing between webs.  A
combinatorial proof of eq.~(\ref{column}) would provide a basis for
generalising the column sum rule, analogous to the sub-row sum rule
introduced in ref.~\cite{Gardi:2011wa}. This would provide a link
between subleading kinematic divergences, and subleading terms in the
$1/N_c$ expansion, as hinted at above.  We will return later in the
paper to the combinatorial problem of determining the vector $s(D)$ for
certain webs, and also see how the weighted column sum rule can be
obtained combinatorially in particular cases.\\

In this section, we have reviewed a number of ideas relating to web
mixing matrices, which pave the way for further combinatorial study of
their general properties. In the following section, we begin such an
investigation, by examining the rank of web-mixing matrices in
particular cases.

\section{The rank of web-mixing matrices}
\label{sec:rank}
Our aim in this section is provide a simple formula for the rank of a
given class of web-mixing matrices, and in doing so to provide a
systematic classification of webs that is convenient for further
study. As explained in the previous section, the rank of a web-mixing
matrix has a simple physical meaning: it dictates how many independent
connected colour factors a given web contributes to in the exponent of
a Wilson line correlator. The general programme for calculating this
exponent at a given order in perturbation theory is as
follows~\cite{Gardi:2013ita}:
\begin{enumerate}
\item Pick a basis of connected colour factors, sufficient to span the
  space of all possible colour factors. This basis is not unique
  e.g. an alternative choice of colour factors can be generated by
  applying Jacobi identities. Reference~\cite{Gardi:2013ita}
  introduced an effective vertex formalism for calculating web graphs,
  which gave a convenient prescription for choosing a basis of colour
  factors.
\item Once a basis has been chosen, one must consider all webs at the
  given perturbative order. For each web, one must ascertain which
  connected colour factors it contributes to, and work out what the
  corresponding combinations of kinematic factors are that accompany
  each colour factor. This can be done either using web-mixing
  matrices, whose eigenvectors of unit eigenvalue are in one-to-one
  correspondence with connected colour factors, or by using the
  effective vertex formalism discussed in ref.~\cite{Gardi:2013ita}.
\item The total contribution to a given colour factor can now be
  calculated, by carrying out the kinematic integrals found in the
  previous step. This will be a gauge invariant quantity, consisting
  of contributions from all webs which contribute to the given colour
  factor. We call such a quantity a {\it gauge-invariant web}, and its
  calculation may benefit from the use of cleverly designed gauges
  such as that presented in ref.~\cite{Chien:2011wz}.
\end{enumerate}
A current priority is to calculate the relevant gauge-invariant web
combinations at three-loop order, which will extend the state of the
art of present-day resummation techniques by a further logarithmic
order, leading to significant improvements in the precision of
collider physics predictions. To this end, all required combinations
of kinematic factors have been presented, for a given colour factor
basis, in ref.~\cite{Gardi:2013ita}. Calculation of the kinematic integrals
is also underway~\cite{integrals,integrals2}, and involves the use of techniques
that potentially generalise to higher orders in perturbation
theory. This leads to the exciting possibility that one could
calculate kinematic factors for given web families for any number of
gluons, which strongly motivates seeking an all-order elucidation of
web properties. One of the simpler - and most useful - properties to
consider is the rank of web-mixing matrices. A general understanding
of which connected colour factors come from which webs is a
prerequisite for constructing gauge-invariant webs at a given
perturbative order, and the rank provides a first step in this
regard.\\

In investigating the above questions, it is convenient to classify
webs according to how many Wilson lines they connect, as was already
done in the three-loop analysis of ref.~\cite{Gardi:2013ita}. In general,
webs may connect $m$ lines at $n$-loop order, where $1\leq m\leq
n+1$. The case of $m=1$ consists of pure self-energy like
contributions, which vanish for massless external lines. The case of
$m=2$ consists of diagrams which also enter the calculation of the
traditional (two-loop) cusp anomalous dimension. The convenience of
this classification is that webs can only mix (under gauge
transformations) with other webs that connect the same number of
lines. Thus, an independent basis of colour factors for $m$-line webs
must be worked out for each value of $m$ separately. \\

\begin{figure}
\begin{center}
\scalebox{0.8}{\includegraphics{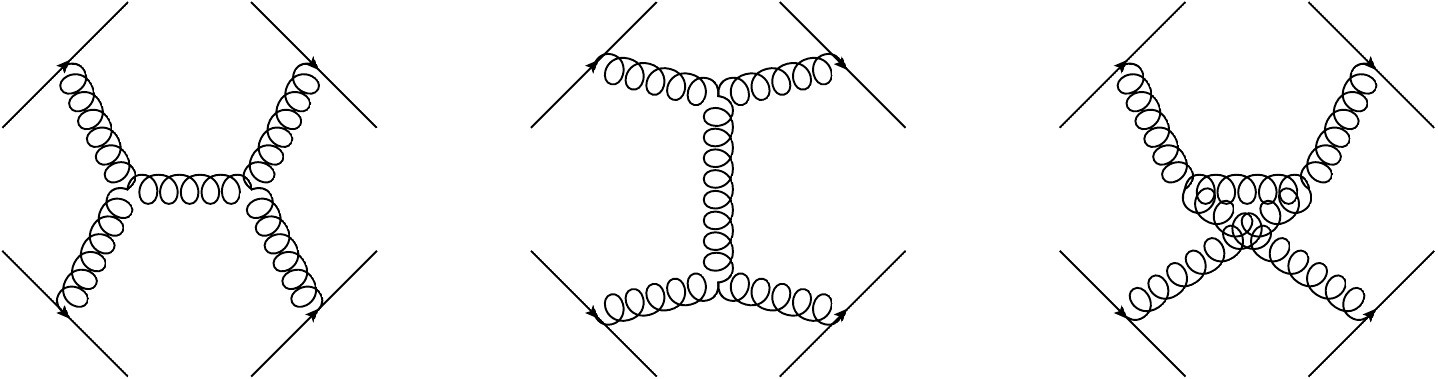}}
\caption{The three possible connected colour factors connecting four
  Wilson lines at three-loop order.}
\label{fig:3loop4}
\end{center}
\end{figure}

Consider first diagrams which connect $n+1$ lines at $n$-loop
order. The possible connected colour factors are shown for the case
$n=3$ in figure~\ref{fig:3loop4}, where we use a notation in which the
Wilson lines are separated from each other for ease of viewing each
gluon diagram (in principle, these Wilson lines all originate
from the same hard interaction point). In the connected colour factor,
each Wilson line emits a single gluon only, as this is the only way in
which the maximal number of lines can be connected.
There are three such colour factors, only two of which are independent
due to their being related by a Jacobi identity. The notation of
figure~\ref{fig:3loop4} makes clear that one can view each connected
colour factor as a 4-point scattering process, where the external
gluons are attached to a Wilson line. Furthermore, each scattering
process is at tree-level, which is a direct consequence of the fact
that $n+1$ lines are connected at $n$-loop order: one can only form
loops by removing a gluon from an external line, such that at most $n$
lines are connected. The number of independent colour factors at
$n$-loop order then amounts to counting the number of independent
colour factors in an $(n+1)$-point tree-level scattering
amplitude. This is known to be $(n-1)!$ (see
e.g. ref.~\cite{DelDuca:1999rs}). \\

\begin{figure}
\begin{center}
\scalebox{0.8}{\includegraphics{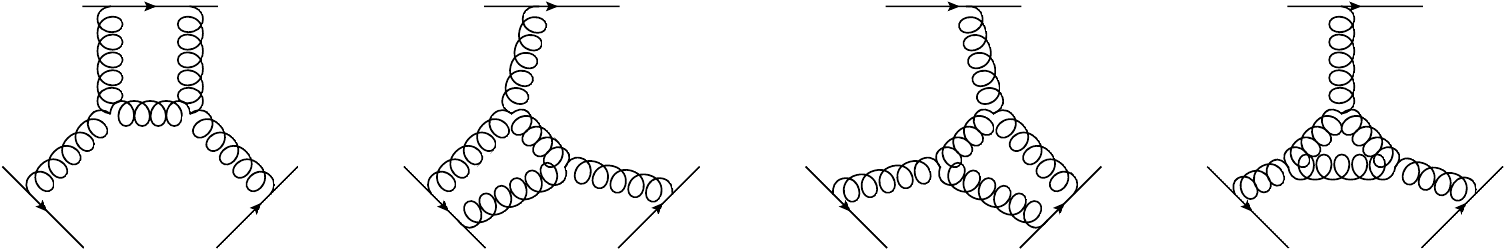}}
\caption{Possible connected colour factors connecting three Wilson
  lines at three-loop order.}
\label{fig:3loop3}
\end{center}
\end{figure}

For webs which link $m<(n+1)$ lines at $n$-loop order, the relevant
connected colour factor graphs may have tree or loop-level
topologies. That is, starting from the case of $n+1$ lines above,
every time one disconnects a given external line from the diagram, one
may either connect to another line more than once, or create a
loop. For example, we show the case of three lines at three-loop order
in figure~\ref{fig:3loop3}.
Four possible connected colour factors are shown, three of which are
tree-level graphs, and one of which is loop-level (and indeed can be
seen to be proportional to the two-loop $Y$ graph)~\footnote{Note that
  the 3-line colour basis presented in ref.~\cite{Gardi:2013ita}
  consists of a superposition of the colour factors in
  figure~\ref{fig:3loop3}.}. For webs connecting progressively fewer
lines at a given order, we can make the following general
observation:\\

{\it Webs which span $1\leq m\leq (n+1)$ Wilson lines at $n$-loop
  order contribute colour factors corresponding to connected 
  gluon graphs, whose topologies contain up to $n+1-m$ loops.}\\

For a given value of $m$, the question arises of how many independent
connected colour factors receive contributions from a given web
(namely, the rank of the web-mixing matrix). It is possible to provide
a simple formula for the case $m=n+1$, namely those webs which connect
$n+1$ lines with $n$ single gluon emissons, such that there are no
three or four-gluon vertices off the Wilson lines (one such example is
the (1,2,2,1) web of figure~\ref{webex}):\\

{\it For webs in which $n+1$ external lines are connected by $n$ single gluon 
exchanges, the rank of the web-mixing matrix is given by
\begin{equation}
r=\prod_{i=1}^{n+1}(n_i-1)!,
\label{rank}
\end{equation}
where $n_i$ is the number of gluon emissions on line $i$.}\\

To show this result, one may use the effective vertex formalism for
webs introduced in ref.~\cite{Gardi:2013ita}, and which was used there
to prove the fully connected colour factor property. As explained in
detail in that paper, products of Wilson line operators may be
rewritten to involve ordinary (rather than path-ordered) exponentials,
thus providing a natural setting in which to use replica arguments
(see refs.~\cite{Laenen:2008gt,Gardi:2010rn}) to prove the connected
colour factor property. This introduces an infinite tower of effective
vertices, where each ``$k$-vertex'' describes the emission of $k\geq
1$ gluons from a single Wilson line. Each such vertex consists of a
superposition of $(k-1)!$ independent fully connected colour factors,
with associated kinematic coefficients. In this language, diagrams
which contribute to the exponent of a Wilson line correlator are
produced by linking $k$-vertices on each Wilson line such that the
overall diagram has a single connected piece. \\

\begin{figure}
\begin{center}
\scalebox{0.8}{\includegraphics{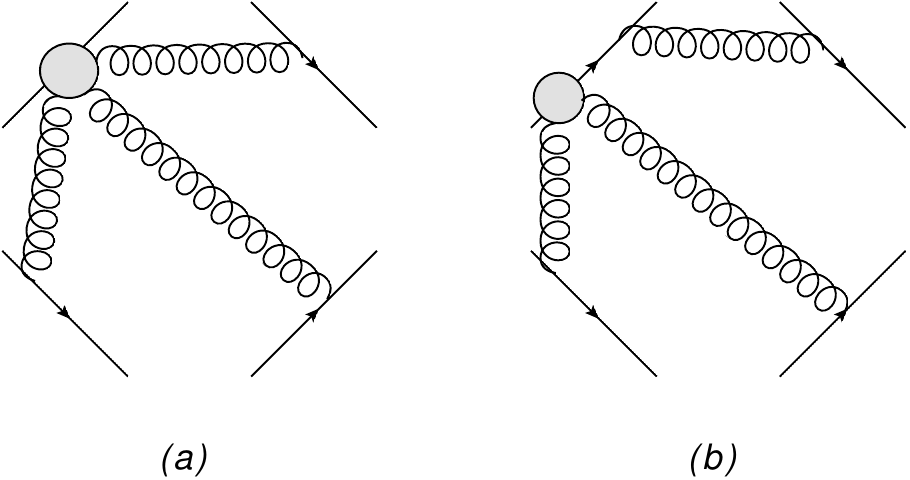}}
\caption{(a) Possible diagram for the (3,1,1,1) web, involving an
  effective 3-vertex on line 1; (b) Configuration involving a 2-vertex
  and a 1-vertex on line 1.}
\label{fig:3111}
\end{center}
\end{figure}

As we have remarked above, the class of webs we are considering gives
rise to connected colour factors in which there is at most a single
gluon emission from each of the $n+1$ external lines. This in turn
means that there can be only one effective vertex on each line. As an
example, consider the case of a (3,1,1,1) web. In
figure~\ref{fig:3111}(a), we show one possible diagram involving a
3-vertex on line 1. This results in a fully connected configuration of
gluons. Alternatively, one may consider a 2-vertex and a 1-vertex on
line 1 - which gives a disconnected gluon diagram, as shown in
figure~\ref{fig:3111}(b). Being disconnected, the latter does not
contribute to the exponent. \\

Having established that for webs connecting $n+1$ lines with $n$
gluons each connected colour factor arises from diagrams in which the
$i^{\rm th}$ Wilson line contains a single $n_i$-vertex, the total
number of independent connected colour factors arising from each web
is simply the product of the number of independent connected colour
factors in each $n_i$-vertex, which (from above) is $(n_i-1)!$. This
gives the result of eq.~(\ref{rank}), as desired. We will see special
cases of this result later in the paper, and also connect it to the
results of ref.~\cite{Dukes:2013wa}.\\

When fewer than $n+1$ lines are linked by gluon exchanges, it is no
longer true that connected colour factor diagrams can only be formed
by having single effective vertices on each line (plenty of examples
can be found in ref.~\cite{Gardi:2013ita}). Nevertheless, it would be
interesting to try to extend the above result, albeit after first
solving for the total number of independent connected colour factors
one expects for $m$-line webs with $m<(n+1)$. This appears to be a
difficult combinatorial problem in general. Note that
ref.~\cite{Dukes:2013wa} derived a result for the $(2,2,\ldots,2)$
web, namely that the rank of the mixing matrix for $n$ gluon exchanges
is given by $(n+1)$. \\

In this section, we have provided a simple formula for the rank of the
mixing matrices for a special class of webs, namely those consisting
of multiple gluon exchanges linking $n+1$ Wilson lines at
$n$-loop order. In fact, for certain of these webs one may go much
further, and explicitly solve for the entire web-mixing matrix. To
this end, and motivated by the pure mathematical study of
ref.~\cite{Dukes:2013wa}, it is useful to relate web diagrams to
partially ordered sets (posets), whose properties we review in the
following section.

\section{Webs and posets}
\label{sec:posets}

In section~\ref{sec:review}, we have reviewed what is known about web
mixing matrices, with a particular emphasis on how they can be
obtained combinatorially via overlap functions. In this section, we
introduce mathematical ideas regarding partially ordered sets
(posets). These were used in ref.~\cite{Dukes:2013wa} to study web
mixing matrices, and will be reviewed here as a precursor for
presenting the results of that paper. Furthermore, in what follows we
will be able to use these poset ideas to obtain an explicit solution
for the web-mixing matrix for $(1,2,2,\ldots,2,1)$ webs, and
investigate the weighted column sum rule in this context. Given that
posets are not necessarily well-known to a physics audience, we will
summarise relevant facts about them here in order to make our
presentation self-contained. A pedagogical introduction may be found
in e.g. ref.~\cite{Mazur}.\\

A poset is a set of objects $X$ endowed with an ordering operation $\leq$ 
possessing the following properties:
\begin{enumerate}
\item {\it Reflexivity}. If $x\in X$ then $x\leq x$.
\item {\it Antisymmetry}. If $x\leq y$ and $y\leq x$ then $x=y$.
\item {\it Transitivity}. If $x\leq y$ and $y\leq z$, $x\leq z$.
\end{enumerate}
This ordering operation applies to pairs of elements in the set $X$,
but such that not all pairs are necessarily ordered with respect to
each other (hence the ``partially'' in ``partially ordered set''). A
familiar example of a poset is the set of integers, where $\leq$ has
its usual meaning of ``less than or equal to''. In this case, however,
the set is fully ordered, with each element being ordered with respect
to every other element.\\

\begin{figure}
\begin{center}
\scalebox{0.8}{\includegraphics{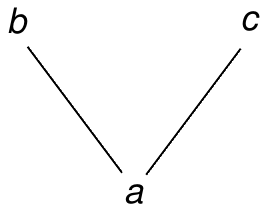}}
\caption{Example Hasse diagram for the poset $\{a,b,c\}$, with $a\leq c$ 
and $a\leq c$.}
\label{posetex}
\end{center}
\end{figure}

The structure of a given poset can be represented on a {\it Hasse
  diagram}.  In such diagrams, each element of $X$ is represented by a
symbol, and a line drawn between distinct elements $x$ and $y$ only if
$x\leq y$ and $\{z \in X: x\leq z\leq y\}=\{x,y\}$. Then $x$ is drawn
below $y$ to indicate the ordering. An example is shown in
figure~\ref{posetex}, for the set of elements $\{a,b,c\}$, such that
$a\leq b$, $a\leq c$ but no ordering is implied between elements $b$
and $c$.\\

\begin{figure}
\begin{center}
\scalebox{0.7}{\includegraphics{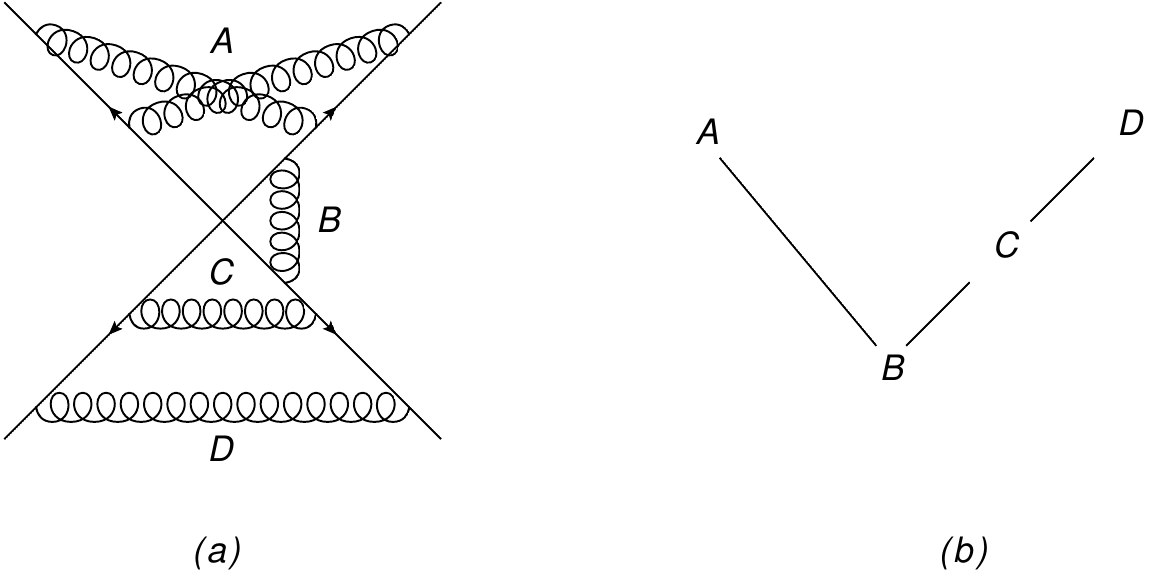}}
\caption{(a) Example web diagram; (b) Hasse diagram corresponding to the
poset generated by the diagram in (a).}
\label{posetex2}
\end{center}
\end{figure}

Returning to the web diagrams introduced in section~\ref{sec:review},
every web diagram $D$ can be identified with a poset. The elements of
the poset are the various irreducible subdiagrams that occur in the
diagram $D$. The ordering operation $\leq$ is defined as follows:
given two irreducible subdiagrams $x$ and $y$, $x\leq y$ if $x$ lies
closer to the origin than $y$, in the sense that one cannot pull $y$
into the origin without also pulling in $x$. As an example, consider
the diagram shown in figure~\ref{posetex2}(a).  In this diagram, the
two gluons labelled $A$ form an irreducible subdiagram: the fact that
they are crossed means that one cannot shrink either gluon to the
origin independently of the other one. The single gluon exchanges
labelled $B$, $C$ and $D$ are also irreducible subdiagrams. Then the
structure of the diagram is encoded by the information that $B$ must
be shrunk before $A$, $C$ or $D$, and $C$ must be shrunk before
$D$. This corresponds to the Hasse diagram shown in
figure~\ref{posetex2}(b). \\

Each diagram $D\in W$, for some web $W$, corresponds to a different
Hasse diagram. One may then consider the set of Hasse diagrams instead
of the original Feynman diagrams. Note that each Hasse diagram in a
given web does not necessarily have the same number of vertices. This
is because irreducible subgraphs from one web diagram may combine to
make a higher-order irreducible piece in a different web diagram. An
example is the crossed gluon pair in figure~\ref{posetex2}(a): there
is another diagram in the set in which the two gluons are uncrossed,
and thus appear as two separate vertices in the corresponding Hasse
diagram. It should be clear from these remarks that all web diagrams
can be reprepresented in terms of posets, including those with three
and four-gluon vertices off the Wilson lines. All of the diagrams in a
given web have Hasse diagrams whose vertices are elements of the set
of all possible irreducible subdiagrams.\\

The question of whether a given web diagram can be made out of
partitions of another web diagram can in principle be rephrased in
terms of the Hasse diagrams of the posets of the two diagrams (we will
see examples of this later in the paper).  However, note that
partitions of a web diagram do not necessarily correspond to simple
colourings of the vertices of its Hasse diagram. An example is again
provided by the crossed gluon pair in
figure~\ref{posetex2}(a). Partitions exist in which the two gluons are
painted different colours. Multiplying together the elements of such a
partition cannot produce a crossed gluon pair, and thus can only
produce diagrams in which $A$ is replaced by two vertices, each
representing a single gluon exchange. \\

For webs in which all irreducible subdiagrams are connected, there
indeed exists a well-defined correspondence between colourings of the
web diagrams $D\in W$, and colourings of the vertices of the
corresponding Hasse diagrams.  In this paper, we will be concerned
with special cases of this, in which: (a) web diagrams consist
exclusively of irreducible subdiagrams containing multiple gluon
exchanges only; (b) any two Wilson lines have at most one gluon
exchange between them. For the special cases we consider, the Hasse
diagrams have a simple form, allowing a full solution of the relevant
web-mixing matrices. \\

\begin{figure}
\begin{center}
\scalebox{0.8}{\includegraphics{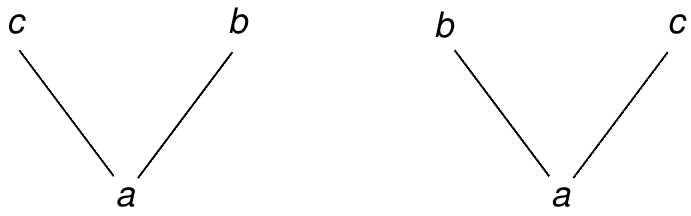}}
\caption{The two linear extensions of the poset shown in figure~\ref{posetex}.}
\label{fig:linex}
\end{center}
\end{figure}

As well as the above notions relating to posets, we will also need the
concept of a {\it linear extension} in what follows. A linear
extension of a given poset $P=(X,\leq)$ is defined as a permutation of
the elements of $X$, such that all pairwise ordering relations remain
satisfied. A simple example is provided by the poset whose Hasse
diagram is shown in figure~\ref{posetex}.  This is specified by the
set $\{a,b,c\}$ and the relations $a\leq b$, $a\leq c$. There are in
fact two Hasse diagrams consistent with this ordering, shown in
figure~\ref{fig:linex}, and thus the number of linear extensions for
this poset is equal to~2. In general the classification of linear
extensions is complicated - it is at least known that the problem of
counting the linear extensions of a poset is
$P$-complete~\cite{Brightwell}.
Given a poset $P$, we denote the set of linear extensions of $P$ by
${\cal L}(P)$. It can then be proven quite generally that the diagonal
elements of any web-mixing matrix are given by ref.~\cite{Dukes:2013wa}
\begin{equation}
R_{DD}=\sum_{\pi\in{\cal L}(P)}\frac{(-1)^{{\rm des}(\pi)}}{p
{p-1 \choose {\rm des}(\pi)}}.
\label{Rdiag}
\end{equation}
Here $\pi$ is a given linear extension of the poset corresponding to
diagram $D$, and $p$ the number of members in the set (i.e. the number
of vertices in the corresponding Hasse diagram). As stated above,
$\pi$ corresponds to a permutation of the elements of the set, and one
may label this by a permutation of $[12\ldots p]$. For example, we
may label the two Hasse diagrams in figure~\ref{fig:linex} (each
corresponding to a linear extension of the same poset) by [123] and
[132], where we have identified $(a,b,c)=(1,2,3)$. Given a
permutation from the original poset (in this case [123]) to one of its
linear extensions, we may count the number of {\it descents} in this
permutation. That is, the number of consecutive numbers $ij$ in the
permutation such that $j<i$. This is denoted by ${\rm des}(\pi)$ in
eq.~(\ref{Rdiag}), and in the above example one has ${\rm des}[123]=0$
and ${\rm des}[132]=1$. \\

Note that the process of sequentially shrinking subdiagrams to the
origin, discussed above from the combinatorial perspective, directly
relates to the ultraviolet subdivergences present in these
diagrams. This was observed already in
refs.~\cite{Mitov:2010rp,Gardi:2010rn} and analysed in detail in
ref.~\cite{Gardi:2011yz}. We will return to this point in
section~\ref{sec:colsum}.\\

Having introduced various concepts relating to posets, and their
relationship to web diagrams, we next consider the application of
these ideas to specific web families. This is the subject of the
following section.

\section{General solutions for web-mixing matrices}
\label{sec:results}
In the previous section, we presented the notion of a partially
ordered set, and explained how this concept relates to web
diagrams. Methods from the theory of posets were used
in ref.~\cite{Dukes:2013wa} to prove general properties of web-mixing
matrices for two special cases, namely $(1,1,\ldots,1,n)$ and
$(1,2,2,\ldots,2,1)$ webs, where we have used the notation introduced
in section~\ref{sec:review}. An expression for the rank of mixing
matrices for $(2,2,\ldots 2)$ webs was also given. Our aim here is not
to reproduce the proofs contained in ref.~\cite{Dukes:2013wa}, to which we
refer the reader for all relevant details. Rather, our goal in the
present study is to use the poset ideas from the previous section to
go further than ref.~\cite{Dukes:2013wa} in solving for the web-mixing
matrices of interest explicitly, to all orders in perturbation theory.\\

\subsection{The $(1,2,2,\ldots,2,1)$ web}
\begin{figure}
\begin{center}
\scalebox{0.8}{\includegraphics{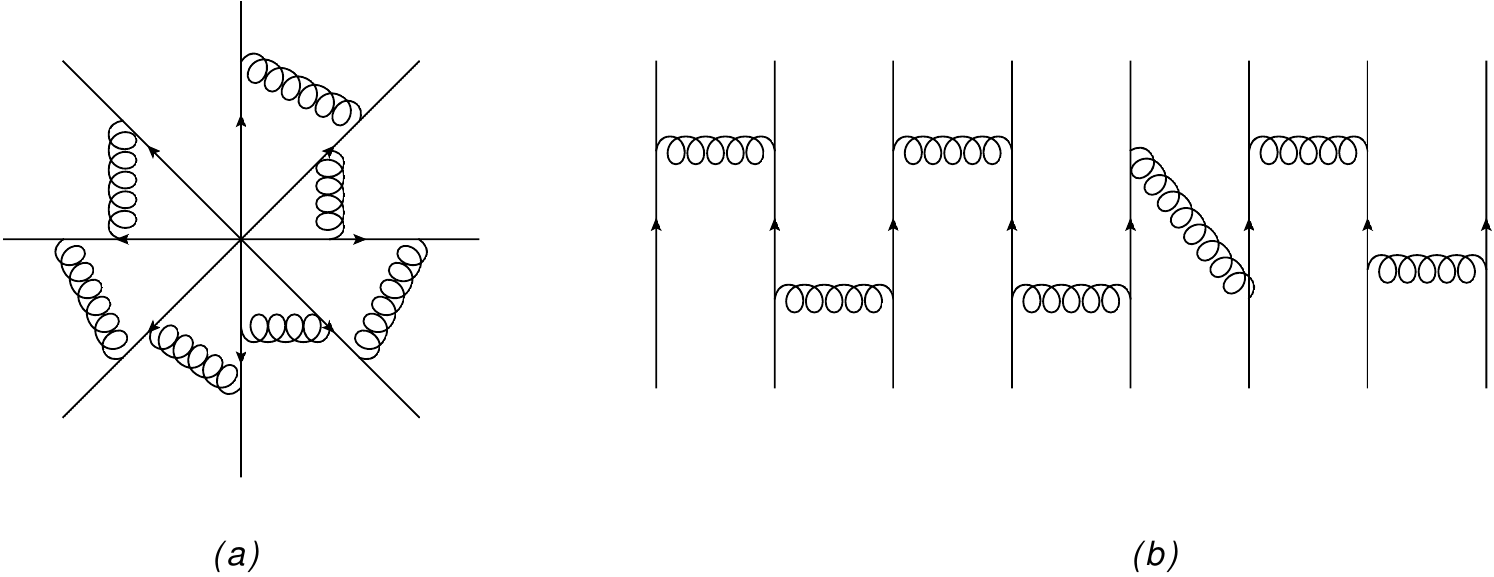}}
\caption{(a) An example (1,2,2,2,2,2,2,1) web; (b) the same diagram 
``unwrapped'', such that the hard interaction corresponds to a horizontal line
at the bottom of the diagram.}
\label{fig:web8}
\end{center}
\end{figure}

In this section we consider webs in which at least one, and at most
two attachments occur on any given leg. One may choose to label the
external lines such that the pattern of exchanges is labelled by
$(1,2,\ldots,2,1)$. This notation does not fully specify the family of
webs of interest: we will further assume that all consecutive lines
are connected by gluon exchanges. An example is shown in
figure~\ref{fig:web8}(a), for the case of seven gluons connecting
eight Wilson lines, and where all external lines are shown emanating
from the same hard interaction vertex. It is also convenient to use an
alternative representation of these diagrams, in which all external
lines are shown parallel to each other.
That is, one may redraw the diagram of figure~\ref{fig:web8}(a) as shown in
figure~\ref{fig:web8}(b). Shrinking gluons to the origin in the former amounts
to sinking gluons in the latter, and the hard interaction can be thought of as
a horizontal line at the bottom of the diagram. \\

The $(1,2,2,\ldots 1)$ web family is of physical interest because it
connects $n+1$ external lines with $n$ gluons. This is the maximum
possible number of external lines which can be entangled at a given
loop order, and corresponds to the class of webs whose rank was
discussed in section~\ref{sec:rank}. A given number of gluon exchanges
$n$ specifies a unique $(1,2,2,\ldots 1)$ web (other similar webs at a
given loop order can be obtained by permuting the external lines). The
number of diagrams in this web is equal to the dimension of the
permutation group which acts on the gluon emission vertices, namely
\begin{displaymath}
{\rm dim}\left[S_1\times \underbrace{S_2\times S_2\ldots S_2}_{\textrm{$n-1$ 
factors}}\times S_1\right]=2^{n-1}.
\end{displaymath}
We collect explicit examples from $n=2$ to $n=4$ in appendix~\ref{app:webs},
together with the relevant mixing matrices. \\

As stated above, this family has been considered in ref.~\cite{Dukes:2013wa}, where 
some general properties of the web-mixing matrices for all $n$ have been 
proven. Of relevance for the current study is that, for these webs, one has
\begin{equation}
{\rm Tr}\left[R_{(1,2,\ldots,2,1)}\right]=1.
\label{TrR1}
\end{equation}
Given the rank is equal to the trace for an idempotent matrix, this
provides a special case of eq.~(\ref{rank}), which dictates that the
mixing matrix of any $(1,2,2,\ldots,2,1)$ web has unit rank. Thus,
there is only one linearly independent row or column. Here we use this
property, together with the structure of the Hasse diagrams
corresponding to each web diagram, to write down a solution for the
web-mixing matrix for any $n$, extending the results of
ref.~\cite{Dukes:2013wa}.\\

\begin{figure}
\begin{center}
\scalebox{0.8}{\includegraphics{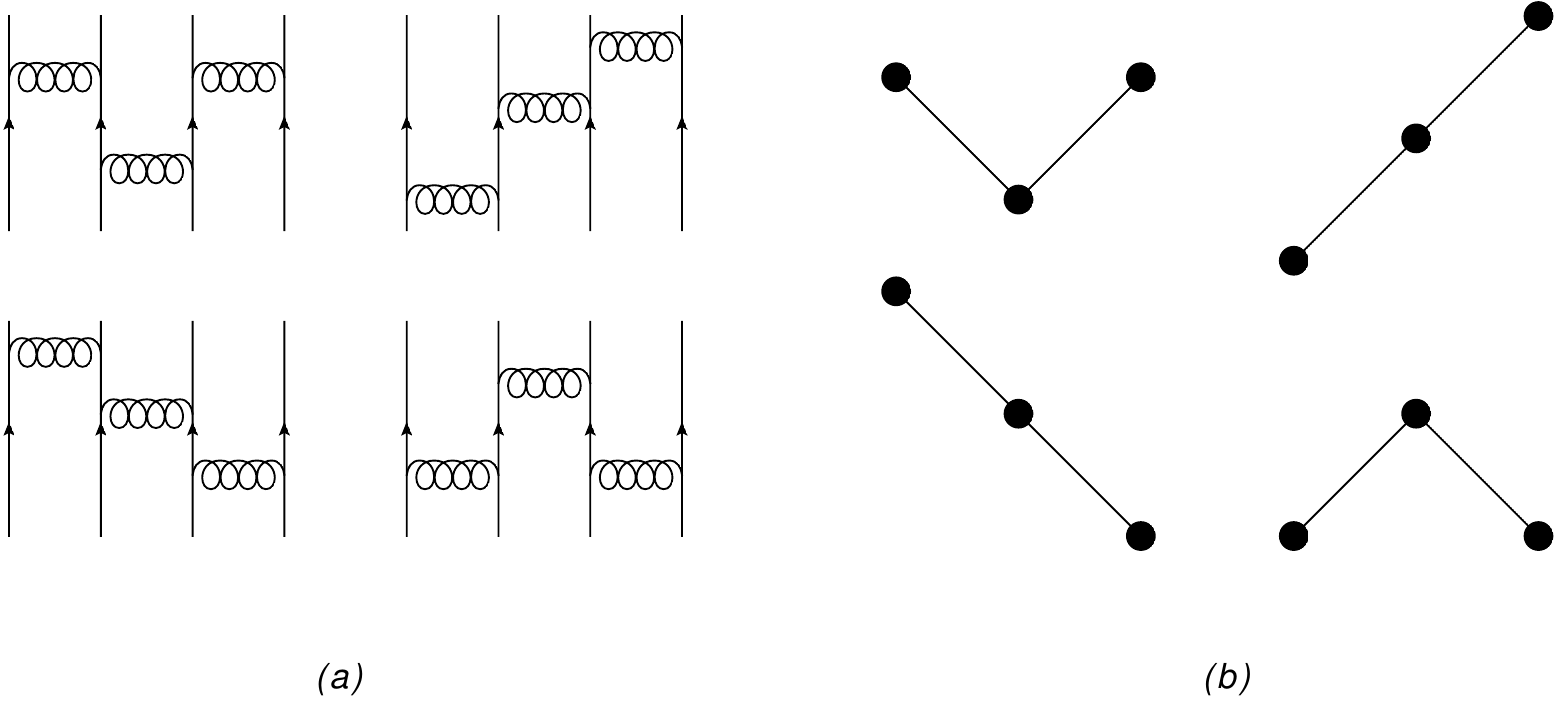}}
\caption{(a) The (1,2,2,1) web redrawn as in figure~\ref{fig:web8}(b); (b) 
corresponding Hasse diagrams.}
\label{poset1221}
\end{center}
\end{figure}

\begin{figure}
\begin{center}
\scalebox{0.8}{\includegraphics{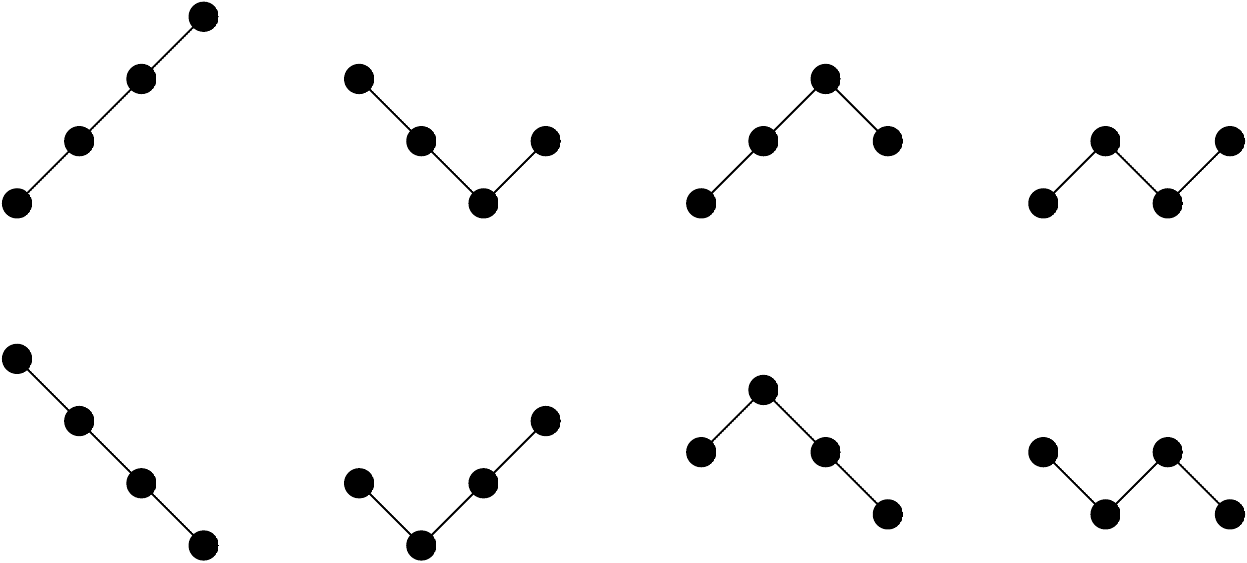}}
\caption{Hasse diagrams for the (1,2,2,2,1) web.}
\label{fig:Hasse4}
\end{center}
\end{figure}

The first thing to do is to relate the diagrams in a
$(1,2,2,,\ldots,2,1)$ web to Hasse diagrams. We can do this by drawing
the diagrams in the form shown in figure~\ref{fig:web8}(b), and then
associating a vertex $\bullet$ with each gluon exchange. This is
possible because all irreducible subdiagrams correspond to a single
gluon exchange. An example is shown in figure~\ref{poset1221} for the
(1,2,2,1) web.
Note that all the Hasse diagrams correspond to a single chain, which
may or may not have kinks in. This is due to the special nature of the
$(1,2,\ldots,2,1)$ web, namely that each gluon exchange has only two
neighbouring gluons (on adjacent external lines), and no other gluons
exchanged between its own pair of external lines. This allows us to
immediately write down the Hasse diagrams for any number of gluons
$n$: one simply draws all possible kink assignments in a chain with
$n-1$ links. As a further example, the case of $n=4$ is shown in
figure~\ref{fig:Hasse4}.\\

\begin{figure}
\begin{center}
\scalebox{0.8}{\includegraphics{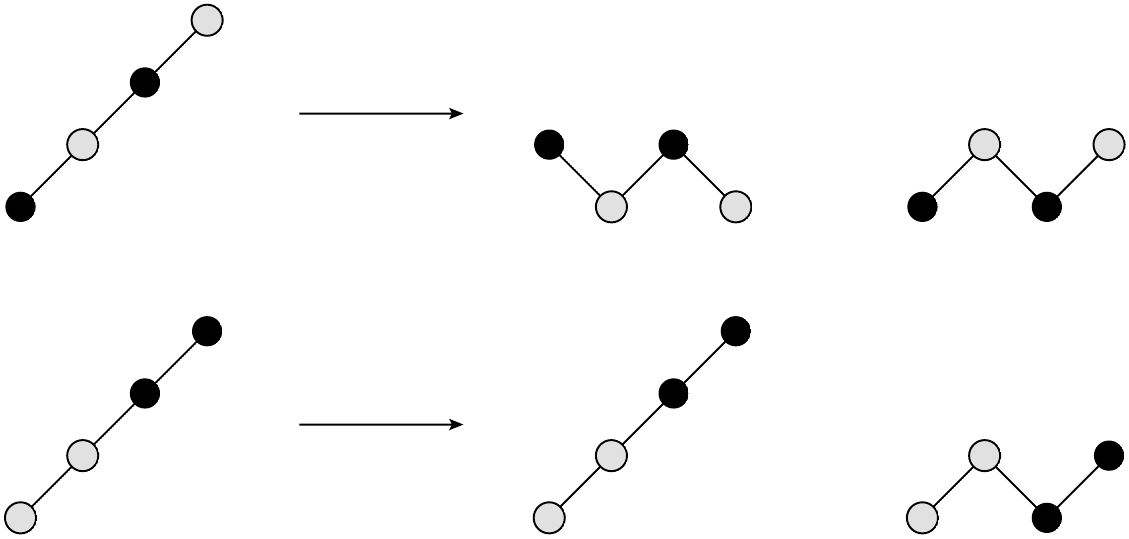}}
\caption{Example 2-partitions of a (1,2,2,2,1) web diagram, and the diagrams
formed by taking products of the partition elements.}
\label{Hasseprod}
\end{center}
\end{figure}

The special form of these Hasse diagrams provides a basis for further
study of the web-mixing matrix, and also the shrinking factors $s(D)$
which enter the weighted column sum rule. In order to investigate the
former, we need to be able to evaluate the overlap functions
introduced here in section~\ref{sec:review}. That is, the number of
ways of forming a diagram in a $(1,2,\ldots,2,1)$ web from a given
partition of some other diagram in the same web. To this end, one may
first see what partitions look like in terms of Hasse diagrams. As
discussed in section~\ref{sec:review}, a partition corresponds to a
colouring of the connected subdiagrams in a given web diagram.  In the
present case, this amounts to a colouring of gluons, namely of the
vertices of the Hasse diagram. A given $m$-partition separates the
Hasse diagram into $m$ groups of vertices. Furthermore, the
multiplication rule for web diagrams easily translates into a
multiplication rule for the elements of a partitioned Hasse
diagram. Examples with $n=4$ are shown in figure~\ref{Hasseprod},
which shows two different 2-partitions of the same web diagram, and
the diagrams obtained by taking products of the partition
elements. One sees that the effect of the colouring is to break links
between consecutive vertices if these have different colours;
otherwise links are left intact. In each product, the grey element of
the partition can be placed above or below the black element, due to
the fact that the product of the related Feynman diagrams corresponds
to placing subgraphs further or nearer to the hard interaction. It is
straightforward to generalise this example to higher numbers of gluons
($n$), and number of colours ($m$).\\

We can now use these ideas to solve explicitly for a particular column
of the web-mixing matrix. We start by considering the Hasse diagram
consisting of a chain with no kinks, such that each vertex is above
the vertex to its left. For the (1,2,2,1) case, this is the diagram in
the upper-right of figure~\ref{poset1221}(b); for $n=4$, this is the
first diagram in figure~\ref{fig:Hasse4}. It is clear that such a
diagram exists for any value of $n$, and we denote this diagram
generically by $\uparrow$. It is then possible to find exactly the
overlap functions $\langle \uparrow|P_D\rangle$, namely the number of
ways of making diagram $\uparrow$ from an arbitrary partition $P_D$ of
diagram $D$ in the same web. One may then, in turn, solve for the
values $R_{D\uparrow}$, namely the entire column of the web matrix
corresponding to the diagram $\uparrow$. \\

The first step is to note that if $\uparrow$ can be made from the partition
$P_D$, there is only one possible way to do this. This is because the various
vertices of the Hasse diagram must be multiplied together in the unique 
sequence which generates the ascending pattern of $\uparrow$. Thus, in general
the overlap function $\langle \uparrow|P_D\rangle$ can only be one or zero.
Furthermore, it is clear that $\uparrow$ cannot be made out of partitions in
which vertices of the same colour are separated by one or more vertices of a 
different colour. An example of this is shown in the upper diagram of 
figure~\ref{Hasseprod}, which has two grey vertices separated by a black
vertex. In multiplying the partition elements together, the two grey vertices
must occur at the same level, and either above or below the black vertices.
This rules out the possibility of obtaining the diagram $\uparrow$, and this
argument clearly generalises to higher numbers of gluons and colours.\\

\begin{figure}
\begin{center}
\scalebox{0.8}{\includegraphics{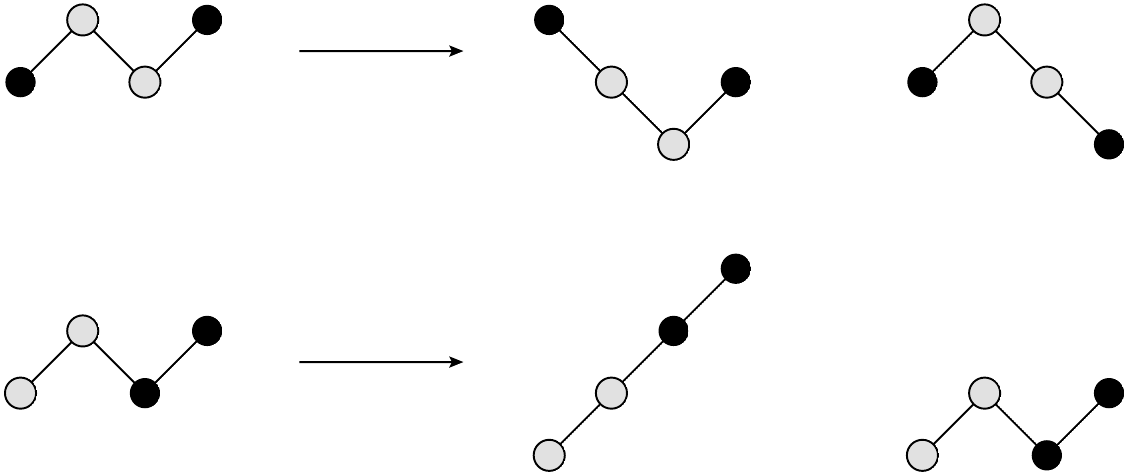}}
\caption{Example 2-partitions of the upper right-most diagram in 
figure~\ref{fig:Hasse4}, and the diagrams formed by taking products of the 
partition elements.}
\label{Hasseprod2}
\end{center}
\end{figure}

It follows that $\langle \uparrow|P_D\rangle$ can only be non-zero for 
partitions in which groups of consecutive vertices are painted the same colour.
Furthermore, in order to make $\uparrow$, none of these groups can contain a
descending link. An example is shown in figure~\ref{Hasseprod2}, in which we
consider 2-partitions of the upper-rightmost diagram in 
figure~\ref{fig:Hasse4}. In the upper partition, a descending link is preserved
by the colouring, which must necessarily survive in the diagrams formed by
taking products of the partition elements. In the lower partition, however, the
descending link is broken by choosing two different colours on either side of
the link. The partition elements then consist solely of groups of ascending 
vertices, and it is then obvious that one can form $\uparrow$ from such a 
partition, as shown explicitly in figure~\ref{Hasseprod2}.\\ 

Let us now consider a diagram $D$ whose Hasse diagram has $r$
descending links, and an $m$-partition of this diagram (equivalently,
a colouring of $D$ with $m$ colours). In order to make $\uparrow$, the
partition must be such that all descending links in the Hasse diagram
are removed. This requires at least $r+1$ colours, such that we must
have $m>r$ for $\langle \uparrow|P_D\rangle$ to be non-zero. After
removing the descending links, any assignment of the remaining colours
amongst the ascending links will give a partition which can reproduce
$\uparrow$. Each of these assignments corresponds to a separate
partition $P_D$, and the number of such partitions amounts to the
number of ways of choosing $m-1-r$ links (the number of remaining
colours) out of $n-1-r$ (the number of ascending links, which is the
total number of links $(n-1)$ minus $r$ descending ones). Thus, one
has
\begin{equation}
\sum_{P_D, n(P_D)=m}\langle \uparrow|P_D\rangle=\Theta(m-r-1){n-1-r\choose m-1-r},
\label{overlapform}
\end{equation} 
where $\Theta(n)$ is the Heaviside function, such that $\Theta(0)=1$. 
Substituting this result into eq.~(\ref{Rdef}), one finds
\begin{align}
R_{D\uparrow}&=\sum_{m=1}^{n}\frac{(-1)^{m-1}}{m}\Theta(m-r-1)
{n-1-r\choose m-1-r}
&=\sum_{m=r+1}^{n}\frac{(-1)^{m-1}}{m}{n-1-r\choose m-1-r},
\label{R121sol}
\end{align}
where we have replaced the sum over individual partitions with a sum over
the number of elements in a partition $m\equiv n(P_D)$. One may now set 
$p=n-1-r$ (the number of ascending links in the Hasse diagram for $D$) to get
\begin{align}
R_{D\uparrow}&=\sum_{m=n-p}^{n}\frac{(-1)^{m-1}}{m}{p\choose m-n+p}\notag\\
&=\sum_{s=0}^p\frac{(-1)^{n-s-1}}{n-s}{p\choose s},
\label{R121sol2}
\end{align}
where we have set $s=n-m$ and used the symmetry property of binomial 
coefficients
\begin{equation}
{p\choose s}={p\choose p-s}
\label{binomsym}
\end{equation}
We may now use the rather quirky identity (see e.g. eq.~(5) of
ref.~\cite{Sury})
\begin{equation}
{n-1\choose p}^{-1}=n\sum_{s=1}^p(-1)^{p-s}
{p\choose s}\frac{1}{n-s}
\label{invid}
\end{equation}
to get 
\begin{equation}
R_{D\uparrow}=\frac{(-1)^{n-p-1}}{n}{n-1\choose p}^{-1}.
\label{R121sol3}
\end{equation}
This is an important formula: it tells us the value of any element of
the column of the web-mixing matrix corresponding to diagram
$\uparrow$, for any number of gluons! Furthermore, the only important
quantity is seen to be the number of ascending links $p$ in the Hasse
diagram of diagram $D$.  Regarding the column as a whole, we can write
a simple expression for the multiplicity $n_p$ of the element given by
eq.~(\ref{R121sol3}). This will simply be the number of Hasse diagrams
with $p$ ascending links. Given that there are $(n-1)$ links in each
Hasse diagram in total, this corresponds to the number of ways of
choosing $p$ links from $(n-1)$, namely
\begin{equation}
n_p={n-1\choose p}.
\label{npsol}
\end{equation}
A consistency check of this result is that the sum over all multiplicities is
\begin{equation}
\sum_{p=0}n_p=\sum_{p=0}{n-1\choose p}
=2^{n-1},
\label{sumnp}
\end{equation}
where we have used a well-known summation formula for binomial coefficients
(see e.g. ref.~\cite{Mazur}). The right-hand side is the dimension of the web-mixing
matrix for the $(1,2,\ldots,2,1)$ web with $n$ gluon exchanges, as it should
be.\\

We can now go further this. The fact that all columns are proportional
to the column $R_{D\uparrow}$ (as follows from eq.~(\ref{TrR1})) means
that we may write the elements of the mixing matrix as
\begin{equation}
R_{DE}=R_{D\uparrow}a_E,
\label{Rij}
\end{equation}
where $a_E$ are coefficients to be determined. A full solution for the
web-mixing matrix now corresponds to finding the latter. To this end,
we may use the result of ref.~\cite{Gardi:2013ita} - namely, that
left-eigenvectors of the web-mixing matrix are in one-to-one
correspondence with connected colour factors. The $(1,2,2,\ldots,2,1)$
web is special in that there is only one such connected colour
factor~\cite{Dukes:2013wa}, and it is straightforward to see how this
arises diagramatically: the only way to form a fully connected colour
factor is to form a commutator of the colour generators on each line
which has two gluon emissions~\footnote{In the effective vertex
  language of ref.~\cite{Gardi:2013ita}, this amounts to having a
  2-vertex on all lines which have two gluon emissions, in line with
  the comments in section~\ref{sec:rank} of the present paper.}. \\

One may show that the left-eigenvector $c_D$ with unit eigenvalue of
the web-mixing matrix is given by:
\begin{equation}
c_D=(-1)^{n-p_D-1},
\label{cEdef}
\end{equation}
where $p_D$ is the number of ascending links in the Hasse diagram
corresponding to $D$. To see this, one may use
eqs.~(\ref{R121sol3}),~(\ref{Rij}) and~(\ref{cEdef}) to find
\begin{equation}
\sum_Dc_DR_{DE}=\sum_D\frac{1}{n}{n-1\choose p_D}^{-1}a_E.
\label{RDEsol1}
\end{equation}
We may convert the sum over diagrams $D$ into a sum over the number of
ascending links $p$ in the corresponding Hasse diagrams, to get
\begin{align}
\sum_D c_DR_{DE}&=\sum_{p=1}^{n}\frac{n_p}{n}{n-1\choose p}^{-1}a_E=\sum_{p=1}^{n}\frac{1}{n}=a_E.
\label{RDEsol2}
\end{align}
In the second step, we have used the result for the multiplicity
factor $n_p$ from eq.~(\ref{npsol}). Equation~(\ref{RDEsol2}) tells us
that $c_D$ of eq.~(\ref{cEdef}) is the left eigenvector we are looking
for, provided
\begin{equation}
c_E=a_E.
\label{cEsol}
\end{equation}
Another way to describe this left eigenvector is as follows:\\

{\it The left eigenvector for the web-mixing matrix for a
  $(1,2,\ldots,2,1)$ web is proportional to $c_E=(-1)^{t(E)}$, where
  $t(E)$ is the number of permutations of gluon emissions that takes
  diagram $\uparrow$ to diagram $E$.\\ }

This is consistent with the fact that the single connected colour
factor for this web has a commutator of gluon generators on each
external line that has two gluon emissions: interchanging any pair of
gluon emissions leads to a minus sign. An example is the (1,2,2,1)
case of figure~\ref{webex}:
\begin{align}
\begin{split}
W^{(1,2,2,1)}&= \sum_{D,D'} {\cal F}(D)  R_{DD'} C(D')
\\
&=\underbrace{\frac16 \bigg({\cal F}(a)-2{\cal F}(b)-2{\cal F}(c)+{\cal F}(d)\bigg)}_{{\cal F}^{(1,2,2,1)}}
\times 
\underbrace{\bigg(C(a)-C(b)-C(c)+C(d)\bigg)}_{C^{(1,2,2,1)}}
\end{split}
\end{align}
where
\begin{align}
\begin{split}
C^{(1,2,2,1)}&= T_1^a (T_2^{ab} T_3^{cb} - T_2^{ba}T_3^{cb} - T_2^{ab} T_3^{bc} + T_2^{ba}T_3^{bc}) T_4^c\\
&= -f^{abc}f^{cbd} T_1^a T_2^e T_3^d T_4^c,\quad \qquad T_i^{ab}\equiv T_i^a\,T_i^b.
\end{split}
\end{align}

Not only has the above analysis determined the left eigenvector of the
web-mixing matrix with unit eigenvalue - it has also fixed the
coefficients $a_E$ in eq.~(\ref{Rij}). We have thus explicitly
obtained a general solution for the web-mixing matrix:\\
\begin{equation}
R_{DE}=\frac{(-1)^{p_D+p_E}}{n}{n-1\choose p_D}^{-1},
\label{R1221res}
\end{equation}
where, as above, $p_D$ is the number of ascending links in the Hasse
diagram of diagram $D$. We may check that this solution satisfies a
number of constraints. Firstly, there is that of idempotence of the
web-mixing matrix. The squared web-mixing matrix is given by
\begin{equation}
\sum_ER_{DE}R_{EF}=\sum_E\left[R_{D\uparrow}c_E\right]\left[R_{E\uparrow}c_F\right],
\label{idempcheck}
\end{equation}
and we may simplify the right-hand side by observing
\begin{equation}
\sum_Ec_ER_{E\uparrow}=c_\uparrow=1.
\label{idempcheck2}
\end{equation}
Inserting this on the right-hand side of eq.~(\ref{idempcheck}) yields
\begin{equation}
\sum_ER_{DE}R_{EF}=R_{D\uparrow}c_F=R_{DF},
\end{equation}
as required. The zero sum row rule is also satisfied, given that
\begin{equation}
\sum_Dc_D=0,
\label{zerosumcheck}
\end{equation}
as follows from eq.~(\ref{cEdef}). The reader may easily verify that
the general solution reproduces the mixing matrices for the cases
$n=2$, $n=3$ and $n=4$, which are collected in
appendix~\ref{app:webs}. For higher values of $n$, we may also compare
eq.~(\ref{R1221res}) with the general combinatorial formula for
diagonal elements of web-mixing matrices, presented in
eq.~(\ref{Rdiag}). A precise comparison at all orders requires a more
involved understanding of linear extensions of the kinked chain than
has been presented here (although we return to this point in
section~\ref{sec:colsum}). Nevertheless, one may note that
eq.~(\ref{Rdiag}) also involves the inverse binomial coefficients.\\

One may also check that our solution satisfies the weighted column sum rule
of eq.~(\ref{column}). However, this is not so straightforward given that one
must first elucidate the structure of the $s(D)$ values for each web. We 
postpone a discussion of this point to section~\ref{sec:colsum}. \\

As an example of the power of the above solution, we can present an
explicit solution for the only surviving combination of kinematic
factors that survives from the $(1,2,2,\ldots,2,1)$ web. From
eqs.~(\ref{Wform}) and~(\ref{Rij}), we see that the contribution of
each web has the form
\begin{align}
W^{(1,2,2,\ldots,2,1)}&=\sum_{DE}{\cal F}_D\,R_{DE}\,C_E\notag\\
&=\left({\cal F}_D R_{D\uparrow}\right)\,(c_E C_E),
\label{kinfac121}
\end{align}
where ${\cal F}_D$ and $C_D$ are the vectors of kinematic and
colour parts. Using eq.~(\ref{R121sol3}) for $R_{D\uparrow}$, the
kinematic combination is given by
\begin{equation}
\sum_{i=1}R_{D\uparrow}{\cal F}_D,\quad R_{D\uparrow}=\frac{(-1)^{n-p_D-1}}{n}{n-1\choose p_D}^{-1}.
\label{eDdef}
\end{equation}
It is useful to clarify the above formula by collecting
the first few non-trivial examples. Firstly there is the $n=2$ case, namely
the web shown in figure~\ref{fig:2loop}. The surviving kinematic combination
is
\begin{equation}
{\cal F}^{(1,2,1)}=\frac{1}{2}\left[{\cal F}(a)-{\cal F}(b)\right],
\label{kin121}
\end{equation}
using the labels from the figure. The $n=3$ case is shown in 
figure~\ref{webex}, and using the labels from the figure one has
\begin{equation}
{\cal F}^{(1,2,2,1)}=\frac{1}{6}\left[{\cal F}(a)-2{\cal F}(b)-2{\cal F}(c)+{\cal F}(d)\right].
\label{kin1221}
\end{equation}
The $n=4$ case is shown in figure~\ref{12221fig}, and contributes the kinematic
combination
\begin{align}
{\cal F}^{(1,2,2,2,1)}&=\frac{1}{24}\Big[6\big({\cal F}[[1],[2,1],[3,2],[4,3],[4]]-{\cal F}[[1],[1,2],[2,3],[3,4],[4]]\big)\notag\\
&\quad\quad-2\big({\cal F}[[1],[1,2],[2,3],[4,3],[4]]+{\cal F}[[1],[2,1],[2,3],[3,4],[4]]\notag\\
&\quad\quad+{\cal F}[[1],[1,2],[3,2],[3,4],[4]]-{\cal F}[[1],[1,2],[3,2],[4,3],[4]]\notag\\
&\quad\quad-{\cal F}[[1],[2,1],[3,2],[3,4],[4]]-{\cal F}[[1],[2,1],[2,3],[4,3],[4]]\big)\Big].
\label{kin12221}
\end{align}
As will be detailed elsewhere~\cite{integrals}, it is possible to
develop systematic methods for carrying out the integration of the
kinematic factors for a general number of gluon exchanges. Combined
with the above results, this would provide a complete solution of this
web family to all orders in perturbation theory, and strongly
motivates the investigation of more general webs.

\subsection{The $(1,1,1,\ldots,1,n)$ web}
\label{sec:sol1n}
In the previous section, we have discussed the $(1,2,2,\ldots,2,1)$ web at
length, and given a full solution for the web-mixing matrix for an 
arbitrary number of gluon exchanges. In this section, we present a solution
for a different web family, namely that of $(1,1,1,\ldots,n)$. This is in fact
a shorter exercise than the previous case, as most of the necessary results
have already been derived in ref.~\cite{Dukes:2013wa}. \\

\begin{figure}
\begin{center}
\scalebox{0.8}{\includegraphics{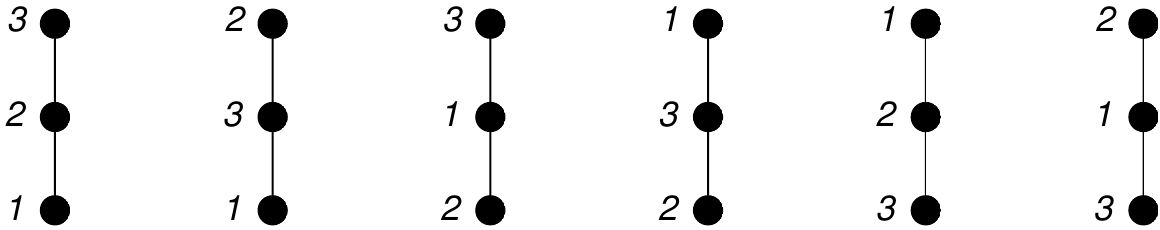}}
\caption{Hasse diagrams for the (1,1,1,3) web.}
\label{Hasse1n}
\end{center}
\end{figure}

The Hasse diagrams for the $(1,1,\ldots,1,n)$ case consist of $n!$ vertical
chains, each corresponding to a permutation of the gluon emissions on the line
which has $n$ gluon emissions. An example is shown, for the $n=3$ case, in 
figure~\ref{Hasse1n}. Consequently, one can label each diagram by a permutation
$\pi$ of $[12\ldots n]$. A general element of the web-mixing matrix can then
be denoted $R_{DD'}\equiv R_{\pi\sigma}$, where $\pi$ and $\sigma$ are the 
permutations corresponding to diagrams $D$ and $D'$ respectively.
We then have the following result~\footnote{Note that we have
  rephrased the definition of $r_{\pi\sigma}$ from
  ref.~\cite{Dukes:2013wa}, where it was given in terms of descents
  rather than ascents. One is allowed to do this due to the reflection
  property of binomial coefficients.}:\\

{\it The web-mixing matrix element $R_{\pi\sigma}$ is given by 
\begin{equation}
R_{\pi\sigma}=\frac{(-1)^{n-r_{\pi\sigma}-1}}{n}{n-1\choose r_{\pi\sigma}}^{-1},
\label{Rpisigma}
\end{equation}
where $r_{\pi\sigma}$ is the number of ascents in the permutation that takes
$\sigma$ to $\pi$.\\
}

The definition of the latter quantity, and how to find the web-mixing
matrix element, is best clarified by a simple example. To this end, we
consider the web of figure~\ref{1311fig}, whose web-mixing matrix is
shown in eq.~(\ref{1311mat}). The ordering of diagrams shown in the
figure can be labelled by the permutations 123, 132, 312, 321, 213 and
231, where these denote the ordering of the gluons (moving outwards
from the hard interaction) on the Wilson line which has three gluon
emissions. Consider now the (3,2) element of the matrix, corresponding
to $R_{(312),(132)}$. The permutation that takes 312 to 132 has 1
ascent (in other words, one consecutive pair of increasing numbers; to
see this, note that it is the same permutation as that which takes 123
to 213), and thus from eq.~(\ref{Rpisigma}) one expects
\begin{displaymath}
R_{(123),(132)}=\frac{(-1)^{3-1-1}}{3}{3-1\choose 1}^{-1}=-\frac{1}{6},
\end{displaymath}
as indeed is observed. As a second example, consider the (2,6) element
of the matrix, corresponding to $R_{(132,231)}$. The permutation which
takes 132 to 231 would take 123 to 321, and thus has 0 ascents. Thus
one expects
\begin{displaymath}
R_{(123),(132)}=\frac{(-1)^{3-1-0}}{3}{3-1\choose 0}^{-1}=\frac{1}{3},
\end{displaymath}
as observed. \\

Some comments are in order. First, the web-mixing matrices for this
family are symmetric i.e. $R_{\pi\sigma}=R_{\sigma\pi}$. This follows
from the definition of $r_{\pi\sigma}$, and the fact that the number
of ascents in the permutation $\sigma\rightarrow\pi$ is the same as
the number of {\it descents} in the inverse permutation
$\pi\rightarrow\sigma$. That is,
\begin{equation}
r_{\pi\sigma}=(n-1)-r_{\sigma\pi}.
\label{rpisigmaflip}
\end{equation}
Swapping $r_{\pi\sigma}\rightarrow r_{\sigma\pi}$ thus leaves the
right-hand side of eq.~(\ref{Rpisigma}) invariant, via the reflection
property of binomal coefficients. Secondly, there is a remarkable
similarity between the expression of eq.~(\ref{Rpisigma}), and the
formula for the possible values in the $(1,2,2,\ldots,1)$ case,
eq.~(\ref{R1221res}). This is because the Hasse diagrams in the
present case are governed by permutations of $[12.\ldots n]$, where
these permutations can be classified by the number of ascents and
descents, and thus mapped to kinked chain diagrams. The multiplicities
of the various values in this case, however, are different, as we
discuss below.\\

Equation~(\ref{Rpisigma}) has a number of corollaries. Firstly, the
permutation that takes a diagram to itself has $n-1$ ascents (i.e. it
takes $[12\ldots n]$ to itself). Thus, the diagonal elements are given
by
\begin{equation}
R_{\pi\pi}=\frac{1}{n}.
\label{Rpipi}
\end{equation}
This can clearly be observed in the examples in
appendix~\ref{app:webs}.  It is also possible to derive this result
directly from eq.~(\ref{Rdiag}), which gives a general formula for the
diagonal elements of any web-mixing matrix (in our case, the number of
vertices in the Hasse diagram is $p=n$). The poset for each web
diagram has a fully ordered Hasse diagram (e.g. figure~\ref{Hasse1n}),
thus has only one linear extension. Furthermore, the number of
descents in the permutation which takes a diagram to itself is zero,
and thus eq.~(\ref{Rpipi}) must correspond to eq.~(\ref{Rdiag}) with
${\rm des}(\pi)=0$. This is indeed the case.\\

Knowledge of the diagonal elements allows us to obtain a simple solution for
the trace:
\begin{equation}
{\rm Tr}\left[R_{(1,1,\ldots,1,n)}\right]=(n-1)!,
\label{TrR1n}
\end{equation}
which follows from the fact that each diagonal element is $1/n$, and
there are $n!$ of them. Again the trace corresponds to the rank, and
thus one derives a special case of the general result of
eq.~(\ref{rank}).\\

Having solved for the possible elements of the web-mixing matrix, it is also
convenient to note their multiplicities in each row and column. These are given
by the following result~\cite{Dukes:2013wa}:\\

{\it Identifying $r\equiv r_{\pi\sigma}$, the values in a given row or column 
of the web-mixing matrix are given by eq.~(\ref{Rpisigma}) with all possible
values of $r$. The multiplicity of each value is given by the Eulerian 
number ${n \bangle r}$.\\}

Given that the Eulerian numbers may not be generally familiar to the reader, 
we briefly introduce their properties in appendix~\ref{app:eulerian}. As
clarification of their role as multiplicities in web-mixing matrices, consider
the $n=4$ case, whose web-mixing matrix is shown in eq.~(\ref{R11114mat}).
In any given row or column, the possible values are given by 
eq.~(\ref{Rpisigma}) with $r=0,1,2,3$, which gives -1/4, 1/12, -1/12 and 1/4
respectively. Furthermore, these should have multiplicities 1, 11, 11 and 1,
as given by the relevant Eulerian numbers from table~\ref{tab:eulerian}. 
This is indeed observed.\\

Note that the sum of multiplicities in any given row or column is
\begin{equation}
\sum_{r=0}^{n-1} {n \bangle r} =n!,
\label{eulersum}
\end{equation}
where we have used a known identity of Eulerian numbers (in words:
summing over $r$ the number of permutations with $r$ ascents must give
the total number of permutations, which is $n!$). The right-hand side
of eq.~(\ref{eulersum}) is the dimension of the web-mixing matrix, as
it should be. Note that the sum over elements in any given row is
\begin{equation}
\sum_{r=0}^{n-1}\frac{(-1)^{n-r-1}}{n}
{n-1\choose r}^{-1}
{n \bangle r}=0,
\label{R1nrowsum}
\end{equation}
where we have used eq.~(\ref{sumReuler2}). This is an explicit
realisation of the zero sum row rule of eq.~(\ref{zerosum}), which has
been proven to be true fully generally in ref.~\cite{Gardi:2011wa}. A
similar sum also ensures that for the $(1,1,\ldots,1,n)$ web family,
the weighted column sum rule is satisfied. We return to this point in
section~\ref{sec:colsum}.\\

In this section, we have presented solutions for two specific families
of web-mixing matrices, for any number of gluon exchanges,
 extending the results of ref.~\cite{Dukes:2013wa}. Next,
we consider the weighted column sum rule of ref.~\cite{Gardi:2011yz}
in the poset language.

\section{The weighted column sum rule}
\label{sec:colsum}
In this section, we return to a specific combinatorial property of web
matrices, namely the weighted column sum rule of
eq.~(\ref{column}). This was first conjectured in
ref.~\cite{Gardi:2011yz}, and is crucially related to the
renormalisation of the hard interaction vertex at which the external
eikonal lines meet. Based on this relation, it was shown in
ref.~\cite{Gardi:2011yz} that for webs which consist solely of
multiple gluon exchanges, the sum rule must hold. However, the sum
rule appears to be more general than this, and was indeed satisfied in
all examples considered in ref.~\cite{Gardi:2011yz}, including those
webs which contain multiple gluon vertices off the eikonal lines. This
perhaps is not surprising, given the results of
ref.~\cite{Gardi:2013ita}, which show that webs containing multiple
gluon exchanges are intimately linked with webs which have multiple
gluon vertices, by virtue of sharing the same connected colour
factors.  They are also linked by gauge transformations. \\

Given the above remarks, it is useful to study the weighted column sum
rule from a purely combinatorial point of view. Whilst a full
combinatorial proof for arbitrary webs remains elusive, we will
relate some key concepts which arise from the poset language of
section~\ref{sec:posets}, and also examine the sum rule for the
particular cases of $(1,2,\ldots,2,1)$ and $(1,1,\ldots,n)$ webs.  \\

\subsection{$s(D)$ and linear extensions}
\label{sec:sDlinex}
Our first step is to tighten up the combinatorial definition of the
number of ways of shrinking irreducible subdiagrams to the origin in a
given web diagram. This may be stated as follows, for all classes of
web:\\

{\it Let $P$ be the poset corresponding to a given web diagram
  $D$. Then the number of ways of shrinking all irreducible
  subdiagrams to the origin is the number of linear extensions of
  $P$.\\ }

This follows from the definition of a linear extension which, as
discussed in section~\ref{sec:posets}, corresponds to a permutation of
the elements of a poset which preserves the ordering requirements. The
Hasse diagram for a web diagram $D$ containing $m$ irreducible
subdiagrams contains $m$ vertices, one for each subdiagram. Each
possible shrinking sequence must preserve the ordering of these
vertices, and corresponds to a permutation of the vertices of the
Hasse diagram. Thus, each shrinking sequence can be uniquely
identified with a linear extension of the poset of $D$, such that the
total number of shrinking sequences is equal to the number of linear
extensions. As an example, consider the diagram of
figure~\ref{posetex2}(a), whose Hasse diagram is shown in
figure~\ref{posetex2}(b). There are three possible shrinking
sequences, which we may denote by $BACD$, $BCAD$, $BCDA$. This is
indeed the number of linear extensions of the poset.\\

Now we recall the definition of $s(D)$ in
ref.~\cite{Gardi:2011yz}. For diagrams where one can sequentially
shrink each connected subdiagram independently to the vertex
(i.e. which are maximally reducible), $s(D)$ is defined to be the
number of shrinking sequences. For non-maximally-reducible diagrams,
$s(D)=0$. It thus follows from the above discussion that for any
maximally-reducible diagram, $s(D)$ can be identified with the number
of linear extensions of the poset of $D$. \\

Having identified $s(D)$ in poset language, we may examine the
weighted column sum rule for the particular cases of $(1,1,\ldots n)$
and $(1,2,2,\ldots,2,1)$ webs considered above. In both of these
cases, all diagrams are maximally reducible, in that all irreducible
diagrams consist of single gluon exchanges. Thus, $s(D)$ for each web
diagram $D$ in either family can be simply identified with the number
of linear extensions of the poset of $D$.

\subsection{Column sum rule for $(1,1,\ldots,1,n)$ webs}
For $(1,1,\ldots,1,n)$ webs, an explicit solution for the web-mixing
matrix elements has been given in eq.~(\ref{Rpisigma}), where $r$
takes values $0,\ldots n-1$, and the multiplicity of each value in a
given column is given by the Eulerian number ${n \bangle r}$. The
$s(D)$ values for any web in this family are trivial, in that they
always have the form
\begin{equation}
s(D)=\left(1\quad1\quad1\ldots 1\right).
\label{sD1n}
\end{equation}
This is easily seen from the fact that the Hasse diagram for any 
$(1,1,\ldots,1,n)$ web diagram is fully ordered (e.g. figure~\ref{Hasse1n}),
and thus all gluon exchanges can only be shrunk in a single sequence, beginning
with the gluon closest to the hard interaction, and working outwards. 
Contracting the vector of $s(D)$ values with any column of the web-mixing 
matrix then gives
\begin{align}
\sum_D s(D)R_{ED}&=\sum_{r=0}^{n-1}\frac{(-1)^{n-r-1}}{n}
{n-1\choose r}^{-1}{n \bangle r}
,\quad \forall E\notag\\
&=0,
\label{sD1n2}
\end{align}
where we have used eq.~(\ref{sumReuler2}). Note that this is analagous to the
proof of the zero sum row property for these webs, as performed in 
eq.~(\ref{R1nrowsum}). This is due to the fact that the same values occur with
the same multiplicities in any given row or column of the web-mixing matrix,
and also crucially depends on the fact that $s(D)$ is the same for all 
diagrams.\\

The fact that the row and column rules appear to be related for this
web family (a consequence in this special case of symmetry of the
web-mixing matrix) is interesting by itself. The row rule is known to
be related to colour properties of web diagrams, including the $1/N_c$
expansion~\cite{Gardi:2011wa}. On the other hand, the column rule is
related to kinematic information, namely the structure of
singularities~\cite{Gardi:2011yz}. A link between the row and column
rules is then a form of colour-kinematic duality, which may be related
to other such dualities in the literature. It would be interesting to
investigate the possible relationship between the row and column rules
in more detail and for more general webs. Another possible question is
that of whether there is a sub-column rule analagous the sub-row rule
introduced in ref.~\cite{Gardi:2011yz} via an expansion about the
planar limit.\\

Here we have seen explicitly that the weighted column sum rule is
satisfied, for this particular family, at the level of individual
webs. Furthermore, this follows purely from combinatorial
considerations. This is consistent with the fact that webs renormalise
individually, with no mixing between webs. Next, we consider the
$(1,2,\ldots,2,1)$ case. 

\subsection{Column sum rule for $(1,2,\ldots,2,1)$ webs}
This web family makes an interesting counterpoint to that of the
previous section. It is nontrivial in that the $s(D)$ values for each
web are different for different diagrams $D$. This in turn implies
that the weighted column sum rule is different to the zero sum row
rule.\\

An explicit solution for a single column of the web-mixing matrix for 
the $(1,2,\ldots,2,1)$ family (corresponding to diagram $\uparrow$) has been 
given in eq.~(\ref{R121sol3}), where each value occurs with a multiplicity 
given by the binomial coefficient of eq.~(\ref{npsol}). All other columns are 
proportional to this column as a consequence of eq.~(\ref{TrR1}) (which 
together with idempotence implies that all web-mixing matrices for this family
have unit rank). Hence, if the column sum rule is satisfied for the column
$R_{E\uparrow}$, it is satisfied for all columns.\\

As stated in section~\ref{sec:sDlinex}, $s(D)$ for a maximally
reducible web diagram is the number of linear extensions of the poset
corresponding to diagram $D$.  In the present case, this corresponds
to linear extensions of posets whose Hasse diagrams correspond to
kinked chains (see e.g. figure~\ref{fig:Hasse4}).  We collect the
corresponding $s(D)$ values up to nine loop order, together with their
multiplicities, in appendix~\ref{app:sDvals}. As we describe in the
appendix, a closed form solution for these values is not
known. However, one may check the weighted column sum rule in
particular cases, for the column $R_{D\uparrow}$. One must first
identify which $s(D)$ values correspond to which diagrams. The value
of $R_{D\uparrow}$ for each diagram is then given by
eq.~(\ref{R121sol3}). As an example, the $s(D)$ values for the $n=3$
case and corresponding to the diagrams in figure~\ref{poset1221} are
2, 1, 1 and 2. The corresponding elements of $R_{D\uparrow}$ are -1/6,
+1/3, +1/3 and -1/6. One then finds $\sum_D s(D) R_{D\uparrow}=0$ as
required.\\

\begin{figure}
\begin{center}
\scalebox{0.8}{\includegraphics{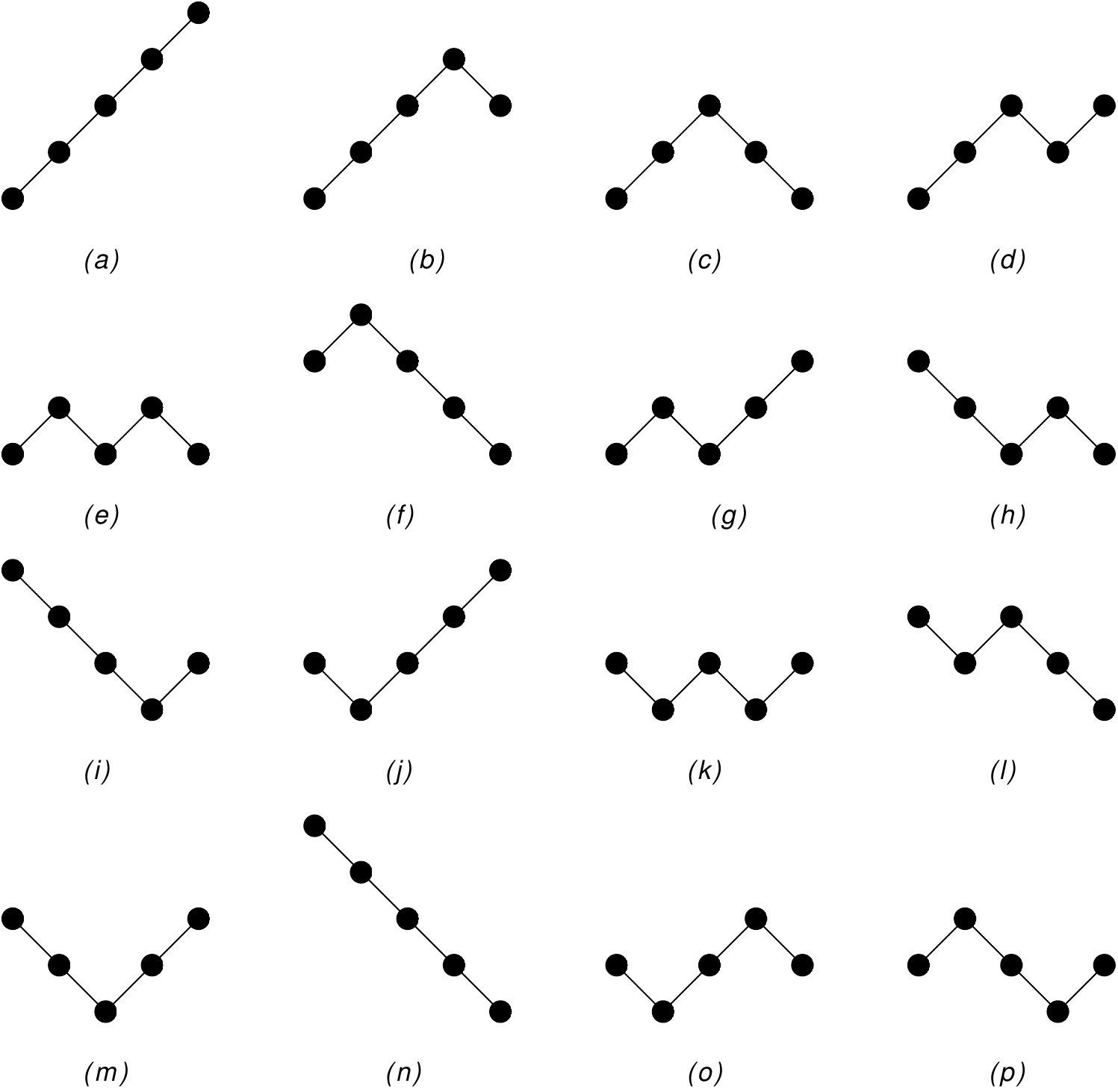}}
\caption{Hasse diagrams for the $(1,2,2,2,2,1)$ web.}
\label{fig:Hassen5}
\end{center}
\end{figure}

As a highly non-trivial example, one may consider also the case $n=5$. The
Hasse diagrams corresponding to the 16 web diagrams are shown in 
figure~\ref{fig:Hassen5}, and the corresponding $s(D)$ factors and 
$R_{D\uparrow}$ values in table~\ref{tab:Hassen5}. Note that we have multiplied
the web-mixing matrices by $5!=120$ in order to make the numbers easier to 
read. 
\begin{table}
\begin{center}
\begin{tabular}{c|c|c||c|c|c}
$D$ & $s(D)$ & $5!R_{D\uparrow}$&$D$ & $s(D)$ & $5!R_{D\uparrow}$\\
\hline
(a) & 1 & 24& (i) & 4 &-6  \\
(b) & 4 & -6& (j) & 4 &-6\\
(c) & 6 & 4 & (k) & 16 & 4\\
(d) & 9 & -6 & (l) &9&-6\\
(e) &16 & 4 & (m) &6&4 \\
(f)& 4& -6 & (n)&1&24\\
(g)& 9 &-6 &(o) & 11&4\\
(h)&9&-6 & (p) & 11 & 4
\end{tabular}
\caption{The factors $s(D)$ for the diagrams of figure~\ref{fig:Hassen5},
together with the corresponding entries of the web-mixing matrix 
$R_{D\uparrow}$, where have multiplied by $5!=120$.}
\label{tab:Hassen5}
\end{center}
\end{table}
Contracting the $s(D)$ values with the web-mixing matrix elements gives
\begin{align*}
\sum_Ds(D)R_{D\uparrow}&\propto \left[1\times24\times2\right]
+\left[4\times(-6)\times4\right]+\left[6\times4\times2\right]\\
&\quad
+\left[9\times(-6)\times4\right]+\left[16\times4\times2\right]+
\left[11\times4\times2\right]\\
&=0,
\end{align*}
as expected if the weighted column sum rule is true. One sees from this
particular example that the nature of the cancellations involved is
highly non-trivial, involving an intimate interplay between the numbers of
eq.~(\ref{R121sol3}), and the linear extensions of the kinked chain.\\

A dramatic simplification occurs for even values of $n$. Consider a
diagram $D$ and its {\it conjugate diagram} $D^*$ whose Hasse diagram
is obtained from that of $D$ by reflection about the horizontal.  The
shrinking factors for these two diagrams are related by
\begin{equation}
s(D)=s(D^*),
\label{SDD*}
\end{equation}
which follows from the fact that the shrinking sequences for diagram $D^*$ can
be obtained by reversing those for $D$. The Hasse diagram of $D^*$ will have $n-1-p$ ascending links if that of $D$ has $p$ ascending links, and thus from eq.~(\ref{cEdef}) one may write
\begin{equation}
R_{D\uparrow}=R_{D^*\downarrow}=(-1)^{n-1}R_{D^*\uparrow}.
\label{Rcolrel3}
\end{equation}
Hence, for even $n$ values, conjugate diagrams (which have the same $s(D)$
values) have column entries in the web-mixing matrix which are equal in 
magnitude but opposite in sign. The weighted column sum rule is then
automatically satisfied, and thus proven for the even $(1,2,2,\ldots,2,1)$
webs.\\

In this section, we have expressed the factors $s(D)$ occuring in the
weighted column sum rule of eq.~(\ref{column}) in poset language,
namely as the number of linear extensions of the poset corresponding
to web diagram $D$, where this is maximally reducible. In the
$(1,1,\ldots,1,n)$ case, we see that the sum rule can indeed be proven
combinatorially. The $(1,2,2,\ldots,2,1)$ case is more intricate, and
involves combinatorial quantities (linear extensions of kinked chain
posets) about which little is currently known. However, we have proven
that the sum rule is satisfied for individual webs, for an even number
of gluon exchanges.  Whilst a formal proof of the weighted column sum
rule (also including other web families) remains elusive, we hope that
the ideas of this section will be useful in the further study of this
property.

\section{Conclusion}
\label{sec:conclude}

The study of infrared singularities in non-Abelian gauge theories
continues to be of interest, not least as this is intimately linked to
the improvement of collider physics predictions through resummation
of large logarithms. In this paper we have continued the development
of a diagrammatic (``web'') approach to soft gluon exponentiation or,
equivalently, correlators of multiple Wilson lines. The web approach
to multiparton scattering identifies closed sets of diagrams, whose
kinematic and colour parts mix according to web-mixing matrices. This
systematic organisation greatly simplifies the calculation of
multiloop contributions, including the structure of subdivergences. 
A full understanding of the exponent necessitates the
development of methods for carrying out the necessary
kinematic integrals -- and these are underway~\cite{integrals,integrals2} --
but also the classification of the mixing
matrices themselves. Here we have focused on the latter. Their pure
combinatorial origin, already identified in ref.~\cite{Gardi:2011wa},
has allowed us to relate the study of web-mixing matrices to that of
partially ordered sets, a connection which was already explored in a
pure mathematical context in ref.~\cite{Dukes:2013wa}.\\

For a special class of webs, namely those consisting of $n+1$ Wilson
lines linked by $n$ gluon exchanges, we have presented a simple
formula for the rank of the mixing matrix. This tells us how many
independent colour factors receive contributions from a given web, and
it is an interesting combinatorial problem in its own right to extend
this to more general webs. \\

Using the poset language of ref.~\cite{Dukes:2013wa}, we have gone
much further than studying the rank in being able to provide explicit
solutions for two particular web families, namely the
$(1,2,2,\ldots,2,1)$ and $(1,1,\ldots,1,n)$ webs. It is interesting to
note that these are extremal cases of the special class mentioned
above - they contribute 1 and $(n-1)!$ independent colour factors
respectively, where the total number of colour factors at a given
order is known to be $(n-1)!$. The formulae for the mixing matrices
involve intriguing combinations of inverse binomial coefficients,
matching the similar behaviour observed for diagonal elements of more
general webs in ref.~\cite{Dukes:2013wa}. Combined with general
methods for carrying out kinematic integrals, details of which will be
presented elsewhere~\cite{integrals,integrals2}, these results are
potentially very powerful, allowing unprecedented insights into
the all-order structure of the {\it exponents} of amplitudes. This
provides strong motivation for generalising the results presented
here, particularly to webs involving three and four-gluon vertices off
the Wilson lines. However, it is also worth stressing that there may
be gauges~\cite{Chien:2011wz} in which the effect of some of these
graphs can be absorbed in webs consisting of multiple gluon exchanges
only, making the web-mixing matrices for the latter more physically
relevant.\\

The poset language also allows some understanding of the weighted
column sum rule for web-mixing matrices presented in
ref.~\cite{Gardi:2011yz}, which implements the cancellation of
subdivergences. While a general combinatorial proof of this property
remains elusive, the poset language seems to provide the right tools
for such a proof. A deeper understanding of the columns of web-mixing
matrices may provide insights into a perceived duality between
subleading colour corrections, and subleading kinematic divergences,
given the zero sub-row sum rule observed in
ref.~\cite{Gardi:2011wa}.\\

Further use of the poset langauge presented here and in
ref.~\cite{Dukes:2013wa} (and possibly other combinatorial methods) can
be used to generalise the results to other web families. Our ultimate
aim is to probe the all-order structure of infrared singularities and
the associated physical consequences, by elucidating where possible
the structure of web-mixing matrices on the one hand, and kinematic
integrals on the other. Progress on both of these fronts is ongoing.

\section*{Acknowledgments}
We thank Einar Steingr\'{i}msson and Jenni Smillie for
discussions. CDW thanks the Higgs Centre for Theoretical Physics for
repeated hospitality. CDW and EG are supported by the UK Science and
Technology Facilities Council (STFC).


\appendix
\section{Examples of web-mixing matrices}
\label{app:webs}
In this appendix, we collect explicit examples of web-mixing matrices for
the $(1,2,2,\ldots 2,1)$ and $(1,1,\ldots,1,n)$ family, where $n$ is the
number of gluon exchanges in each case.
\subsection{$(1,2,2,\ldots2,1)$ webs}
\begin{figure}
\begin{center}
\scalebox{0.8}{\includegraphics{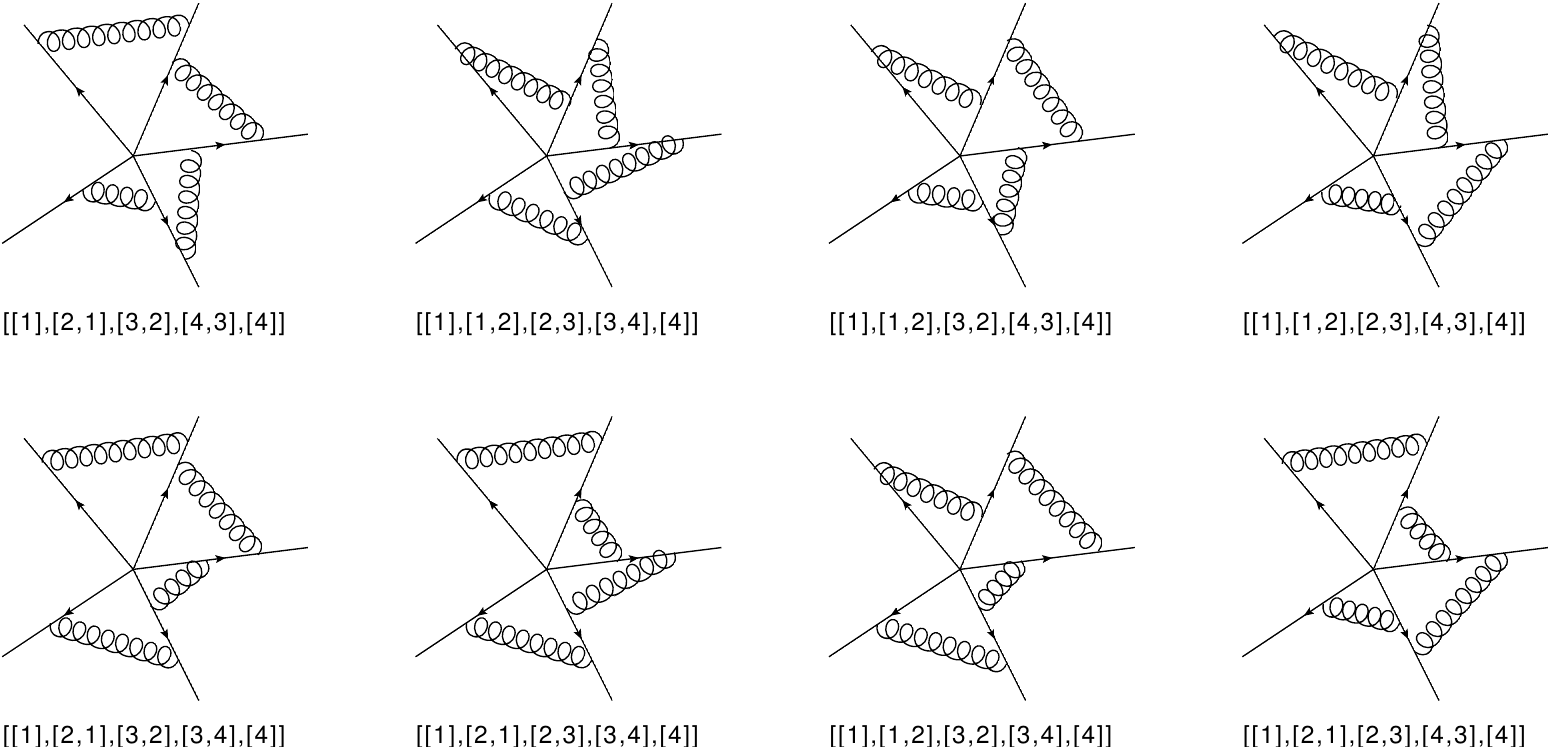}}
\caption{Web whose mixing matrix is given by eq.~(\ref{12221mat}).}
\label{12221fig}
\end{center}
\end{figure}

The web diagrams corresponding to the $n=2$ and $n=3$ cases are shown in 
figures~\ref{fig:2loop} and~\ref{webex} respectively. The corresponding web
mixing matrices are:
\begin{equation}
R_{(1,2,1)}=\frac{1}{2}\left(\begin{array}{rr}1&-1\\-1&1\end{array}\right)
\label{R121}
\end{equation}
and
\begin{equation}
R_{(1,2,2,1)}=\frac{1}{6}\left(\begin{array}{rrrr}1&-1&-1&1\\-2&2&2&-2\\
-2&2&2&-2\\1&-1&-1&1\end{array}\right).
\label{R1221}
\end{equation}
The (1,2,2,2,1) web is shown in figure~\ref{12221fig}, where we label each
diagram according to a notation first introduced in ref.~\cite{Gardi:2010rn},
and in which gluon exchanges associated with the same connected subdiagram
are labelled by integers $i$. The $m$-tuple $[n_1,n_2\ldots,n_m]$ associated
with each external line then labels the gluon emissions on that line, 
working inwards towards the hard interaction vertex. 
In this case the mixing matrix is
\begin{align}
R_{(1,2,2,2,1)}=\frac{1}{24}\left[ \begin {array}{rrrrrrrr} 6&-6&-6&6&-6&6&6&-6
\\ \noalign{\medskip}-6&6&6&-6&6&-6&-6&6\\ \noalign{\medskip}-2&2&2&-2
&2&-2&-2&2\\ \noalign{\medskip}2&-2&-2&2&-2&2&2&-2
\\ \noalign{\medskip}-2&2&2&-2&2&-2&-2&2\\ \noalign{\medskip}2&-2&-2&2
&-2&2&2&-2\\ \noalign{\medskip}2&-2&-2&2&-2&2&2&-2
\\ \noalign{\medskip}-2&2&2&-2&2&-2&-2&2\end {array} \right] 
\label{12221mat},
\end{align}
where the diagrams are ordered as shown in figure~\ref{12221fig}. 

\subsection{$(1,1,1,\ldots1,n)$ webs}
\begin{figure}
\begin{center}
\scalebox{0.8}{\includegraphics{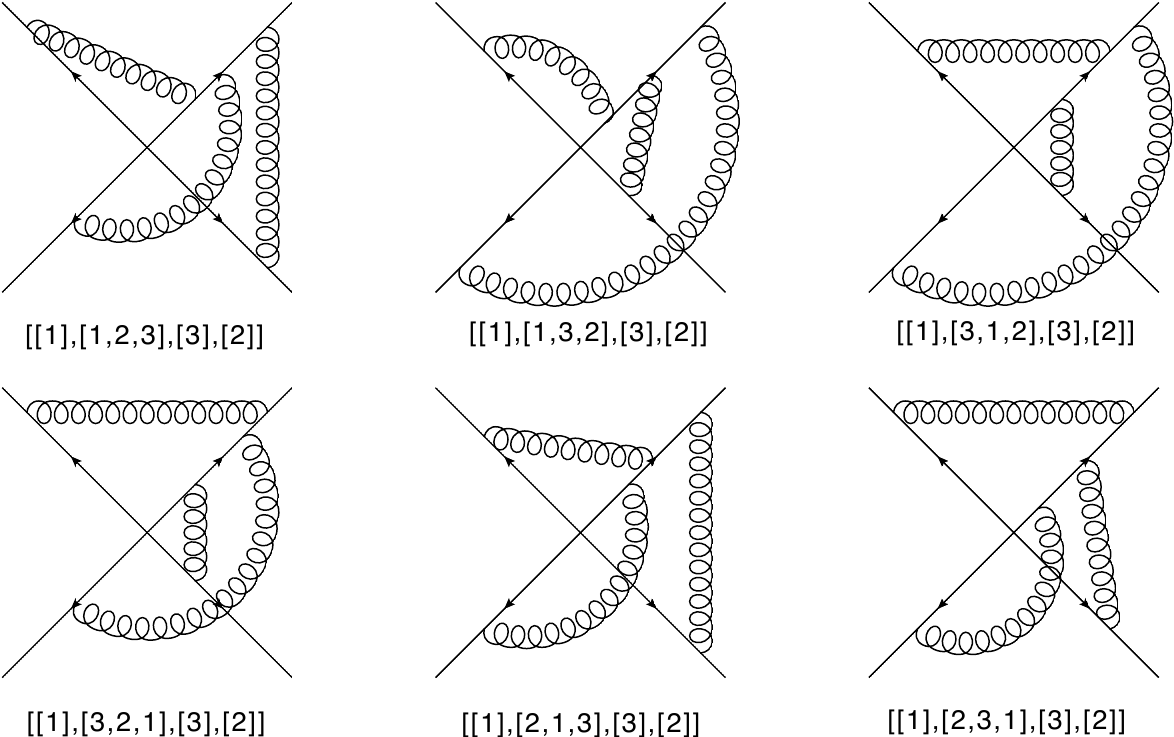}}
\caption{Web whose mixing matrix is given by eq.~(\ref{1311mat}).}
\label{1311fig}
\end{center}
\end{figure}

The $n=2$ case can be obtained by relabelling external lines in the $(1,2,1)$ 
web shown in figure~\ref{fig:2loop}, whose mixing matrix is given in 
eq.~(\ref{R121}). The $n=3$ case is shown in figure~\ref{1311fig}, and the 
mixing matrix is
\begin{align}
R_{(1,1,1,3)}=\frac{1}{6}
\left[ \begin {array}{rrrrrr} 2&-1&-1&2&-1&-1\\ \noalign{\medskip}-1&
2&-1&-1&-1&2\\ \noalign{\medskip}-1&-1&2&-1&2&-1\\ \noalign{\medskip}2
&-1&-1&2&-1&-1\\ \noalign{\medskip}-1&-1&2&-1&2&-1
\\ \noalign{\medskip}-1&2&-1&-1&-1&2\end {array} \right], 
\label{1311mat}
\end{align}
with diagrams ordered as in the figure. \\

For the $n=4$ case $(1,1,1,1,4)$, we do not reproduce the 24 web
diagrams here.  We may label each diagram concisely by a permutation
of (1234), where this denotes the ordering of the gluons on the parton
line with 4 attachments (away from the hard interaction), such that
gluon $i$ joins this line with external line $i$. One then has the
mixing matrix 
\begin{align}
{\tiny\frac{1}{24} \left[ \begin {array}{cccccccccccccccccccccccc} 
6&2&-2&2&-6&-2&2&-2&-
2&2&-2&2&2&-2&-2&-2&2&2&2&-2&-2&2&2&-2\\ \noalign{\medskip}-2&6&-2&2&2
&2&2&2&-6&-2&2&-2&-2&-2&2&-2&2&-2&-2&2&-2&2&2&-2\\ \noalign{\medskip}-
2&-2&6&-6&2&-2&2&2&2&-2&2&2&-2&-2&-2&-2&2&2&-2&2&2&-2&-2&2
\\ \noalign{\medskip}2&2&-6&6&-2&2&-2&-2&-2&2&-2&-2&2&2&2&2&-2&-2&2&-2
&-2&2&2&-2\\ \noalign{\medskip}-6&-2&2&-2&6&2&-2&2&2&-2&2&-2&-2&2&2&2&
-2&-2&-2&2&2&-2&-2&2\\ \noalign{\medskip}-2&2&-2&2&2&6&2&-2&-2&-2&2&2&
-2&2&-2&-2&-2&-6&2&2&-2&2&-2&2\\ \noalign{\medskip}2&2&2&-2&-2&-2&6&-2
&-2&-2&2&2&-2&-2&-2&-6&2&2&2&2&2&-2&2&-2\\ \noalign{\medskip}-2&2&2&-2
&2&-2&-2&6&-2&-2&-2&-2&2&-2&2&2&2&2&-6&2&-2&2&2&-2
\\ \noalign{\medskip}2&-6&2&-2&-2&-2&-2&-2&6&2&-2&2&2&2&-2&2&-2&2&2&-2
&2&-2&-2&2\\ \noalign{\medskip}2&-2&-2&2&-2&-2&-2&-2&2&6&2&-2&-2&-2&2&
2&2&2&2&-6&2&-2&-2&2\\ \noalign{\medskip}-2&2&-2&2&2&2&2&-2&-2&2&6&-2&
-6&-2&2&-2&2&-2&2&-2&2&-2&-2&2\\ \noalign{\medskip}2&-2&2&-2&-2&2&2&-2
&2&-2&-2&6&2&2&-6&-2&-2&-2&2&2&-2&2&-2&2\\ \noalign{\medskip}2&-2&2&-2
&-2&-2&-2&2&2&-2&-6&2&6&2&-2&2&-2&2&-2&2&-2&2&2&-2
\\ \noalign{\medskip}-2&-2&-2&2&2&2&-2&-2&2&-2&-2&-2&2&6&2&2&-6&-2&2&2
&2&-2&2&-2\\ \noalign{\medskip}-2&2&-2&2&2&-2&-2&2&-2&2&2&-6&-2&-2&6&2
&2&2&-2&-2&2&-2&2&-2\\ \noalign{\medskip}-2&-2&-2&2&2&2&-6&2&2&2&-2&-2
&2&2&2&6&-2&-2&-2&-2&-2&2&-2&2\\ \noalign{\medskip}2&2&2&-2&-2&-2&2&2&
-2&2&2&2&-2&-6&-2&-2&6&2&-2&-2&-2&2&-2&2\\ \noalign{\medskip}2&-2&2&-2
&-2&-6&-2&2&2&2&-2&-2&2&-2&2&2&2&6&-2&-2&2&-2&2&-2
\\ \noalign{\medskip}2&-2&-2&2&-2&2&2&-6&2&2&2&2&-2&2&-2&-2&-2&-2&6&-2
&2&-2&-2&2\\ \noalign{\medskip}-2&2&2&-2&2&2&2&2&-2&-6&-2&2&2&2&-2&-2&
-2&-2&-2&6&-2&2&2&-2\\ \noalign{\medskip}-2&-2&2&-2&2&-2&2&-2&2&-2&2&-
2&-2&2&2&-2&-2&2&2&2&6&-6&2&-2\\ \noalign{\medskip}2&2&-2&2&-2&2&-2&2&
-2&2&-2&2&2&-2&-2&2&2&-2&-2&-2&-6&6&-2&2\\ \noalign{\medskip}2&2&-2&2&
-2&-2&2&-2&-2&-2&-2&-2&2&2&2&-2&-2&2&2&2&2&-2&6&-6
\\ \noalign{\medskip}-2&-2&2&-2&2&2&-2&2&2&2&2&2&-2&-2&-2&2&2&-2&-2&-2
&-2&2&-6&6\end {array} \right]},
\label{R11114mat}
\end{align} 

where the ordering of the diagrams is given by
\begin{align*}
&\{2314,1243,3124,4213,4132,3142,4321,4231,3421,4312,2341,4123,1432,2431,3214,
1234,\\
&\quad1342,2413,1324,2134,1423,3241,3412,2143\}.
\end{align*}

\section{Eulerian numbers}
\label{app:eulerian}
In this appendix, we briefly summarise the properties of Eulerian numbers,
which occur in the solution of the web-mixing matrix for $(1,1,\ldots,1,n)$
webs.\\

The {\it Eulerian number} ${n \bangle r}$ counts the number of
permutations of $n$ objects with $r$ ascents (consecutive pairs of
rising numbers), where $0\leq r\leq n-1$. Results for the first few
values of $n$ and $r$ are shown in table~\ref{tab:eulerian}.\\
\begin{table}
\begin{center}
\begin{tabular}{cc|ccccccc}
 & & & &  & r & & &\\
 & & 0 & 1 & 2 & 3 & 4 & 5 & 6\\
\hline
  & 1 & 1 & & & & & &\\
  & 2 & 1 & 1 & & & & &\\
  & 3 & 1 & 4 & 1 & & & &\\
n & 4 & 1 & 11 & 11 & 1 & & &\\
  & 5 & 1 & 26 & 66 & 26 & 1 & &\\
  & 6 & 1 & 57 & 302 & 302 & 57 & 1 &\\
  & 7 & 1 & 120 & 1191 & 2416 & 1191 & 120 & 1
\end{tabular}
\caption{Table of Eulerian numbers.}
\label{tab:eulerian}
\end{center}
\end{table}

A number of identities regarding Eulerian numbers are known. For the present
paper, the following identity is useful:
\begin{equation}
\sum_{r=0}^{n-1}(-1)^r{n-1\choose r}^{-1}
{n \bangle r}=0.
\label{sumReuler2}
\end{equation}

\section{Results for $s(D)$ for $(1,2,2,\ldots,2,1)$ webs}
\label{app:sDvals}
In this appendix, we gather results for the $s(D)$ factors for
$(1,2,2,\ldots 1)$ webs, up to nine loop order. These are given in
table~\ref{tab:sDvals}, which also lists the multiplicities of each
$s(D)$ values. One may check explicitly in each case that the total
number of values is equal to $2^{n-1}$ (the total number of
diagrams). Note that a closed form solution for these numbers (in
terms of known combinatorial functions) is not currently possible. It
is clear from studying table~\ref{tab:sDvals} that there are a number
of interesting relationships. For example, the multiplicities of each
$s(D)$ value are always even. This follows from the fact that the
$s(D)$ value for a given web diagram $D$ must be the same as its
conjugate diagram $D^*$, as well as for the Hasse diagram obtained by
reflecting the Hasse diagram of $D$ about the vertical (by
symmetry). The highest $s(D)$ value in each case corresponds to the
Hasse diagrams which are maximally kinked. Such posets are referred to
as {\it zig-zag posets} or {\it fences} in the mathematical
literature, such that the highest $s(D)$ value for any given $n$ is
the number of linear extensions of the zig-zag poset.
\begin{table}
\begin{center}
\small
\begin{tabular}{| c | c | c || c | c | c || c | c | c || c | c | c |}
\hline
Number of Gluons & $S(D)$ & Multiplicity & & & & & & \\
\hline
n=1 & 1 & 1 &\multirow{27}{*}{}{} & 85 & 4 & & 461  & 4 \\
\cline{1-3}
n=2 & 1 & 2 & & 99 & 4 & & 470 &4 \\
\cline{1-3}

\multirow{2}{*}{}{n=3} & 1 & 2 & & 105 & 4 & & 496 & 4\\
 & 2 & 2 & & 125 &4 & & 512 &4\\
\cline{1-3}

\multirow{3}{*}{}{n=4} & 1 & 2 & & 133 & 4 & & 595 &4 \\
 & 3 & 4 & & 155 & 4 & & 632 &4\\
 & 5 & 2 & & 181 &2 & & 664 &4 \\
\cline{1-3}

\multirow{6}{*}{}{n=5} & 1 & 2 & & 189 & 4 & &685 &2\\
 & 4 & 4 & & 203 & 4 & & 728 & 4\\
 & 6 & 2 & & 217 &4 & & 784 & 4\\
 & 9 & 4 & & 245 &4 & & 785 & 4\\
 & 11 & 2 & & 259 &4 & & 812 & 4\\
 & 16 & 2 & & 315 &8 & & 880 & 2\\
\cline{1-3}

\multirow{9}{*}{}{n=6} & 1 & 2 & & 323 &4 & & 1016 & 4\\
 & 5 & 4 & & 365 &2 & & 1051 & 4\\
 & 10 & 4 & & 407 &4 & & 1099 & 4\\
 & 14 & 4 & & 413 &4 & & 1100 & 4\\
 & 19 & 4 & & 449 &2 & & 1141 & 4\\
 & 26 & 4 & & 477 &4 & & 1168 & 4\\
 & 35 & 4 & & 531 &4 & & 1253 & 4\\
 & 40 & 4 & & 573 &4 & & 1351 & 4\\
 & 61 & 2 & & 589 &2 & & 1421 & 4\\
\cline{1-3}

\multirow{19}{*}{}{n=7} & 1 & 2 & & 643 &4 & & 1456 &2\\
 & 6 & 4 & & 791 &4 & & 1457 & 4\\
 & 15 & 4 & & 875 &4 & & 1513 & 2\\
 & 20 & 6 & & 917 &4 & & 1519 & 4\\
 & 29 & 2 & & 1385 &2 & & 1667 & 4\\
\cline{4-6}
 & 34 & 4 & \multirow{22}{*}{}{n=9}  & 1 &2 & & 1735 &4\\
 & 50 & 4 & & 8 &4 & & 1856 & 4\\
 & 55 & 4 & & 28 &4 & & 1889 &4\\
 & 64 & 4 & & 35 &4 & & 2051 & 4\\
 & 71 & 2 & & 55 &2 & & 2107 & 4\\
 & 78 & 4 & & 56 &4 & & 2144 & 4\\
 & 90 & 2 & & 70 &2 & & 2261 & 4\\
 & 99 & 4 & & 83 &4 & & 2312 & 4\\
 & 111 & 4 & & 125 &4 & & 2590 & 4\\
 & 132 & 2 & & 133 &4 & & 2701 & 2\\
 & 155 & 4 & & 160 &4 & & 2780 & 2\\
 & 169 & 2 & & 161 &4 & & 2890 & 4\\
 & 181 & 4 & & 208 &4 & & 2990 & 4\\
 & 272 & 2 & & 245 &4 & & 3194 & 4\\
\cline{1-3}

\multirow{35}{*}{}{n=8} & 1 & 2 & & 245 & 4 & & 3268 & 4\\
 & 7 & 4 & & 259 &4 & & 3526 & 4\\
 & 21 & 4 & & 295 &4 & & 3736 & 4\\
 & 27 & 4 & & 350 &4  & & 4529 & 4\\
 & 35 & 4 & & 370 &4 & & 4985 & 4\\
 & 41 & 2 & & 379 &4 & & 5095 & 2\\
 & 55 & 4 & & 412 &4 & & 5263 & 4\\
 & 69 & 2 & & 448 &2 & &  7936& 2\\
\hline
\end{tabular}
\normalsize
\caption{$s(D)$ values for the $(1,2,2,\ldots 1)$ web, where $n$ is the 
number of gluon exchanges. Also shown is the multiplicity of each $s(D)$ 
value.}
\label{tab:sDvals}
\end{center}
\end{table}

\bibliography{refs.bib}

\providecommand{\href}[2]{#2}\begingroup\raggedright\begin{thebibliography}{10}

\bibitem{Arefeva:1980zd}
I.~Y. Arefeva, ``{Quantum contour field equations},'' {\em Phys. Lett.} {\bf
  B93} (1980)
347--353.

\bibitem{Polyakov:1980ca}
A.~M. Polyakov, ``{Gauge Fields as Rings of Glue},'' {\em Nucl. Phys.} {\bf
  B164} (1980)
171--188.

\bibitem{Dotsenko:1979wb}
V.~S. Dotsenko and S.~N. Vergeles, ``{Renormalizability of Phase Factors in the
  Nonabelian Gauge Theory},'' {\em Nucl. Phys.} {\bf B169} (1980)
527.

\bibitem{Brandt:1981kf}
R.~A. Brandt, F.~Neri, and M.-a. Sato, ``{Renormalization of Loop Functions for
  All Loops},'' {\em Phys. Rev.} {\bf D24} (1981)
879.

\bibitem{Korchemsky:1985xj}
G.~P. Korchemsky and A.~V. Radyushkin, ``Loop space formalism and
  renormalization group for the infrared asymptotics of {QCD},'' {\em Phys.
  Lett.} {\bf B171} (1986)
459--467.

\bibitem{Ivanov:1985np}
S.~Ivanov, G.~Korchemsky, and A.~Radyushkin, ``Infrared asymptotics of
  perturbative {QCD}: {C}ontour gauges,'' {\em Yad. Fiz.} {\bf 44} (1986)
230--240.

\bibitem{Korchemsky:1985xu}
G.~Korchemsky and A.~Radyushkin, ``Infrared asymptotics of perturbative {QCD}:
  {R}enormalization properties of the wilson loops in higher orders of
  perturbation theory,'' {\em Sov. J. Nucl. Phys.} {\bf 44} (1986)
877.

\bibitem{Korchemsky:1985ts}
G.~Korchemsky and A.~Radyushkin, ``Infrared asymptotics of perturbative {QCD}.
  {Q}uark and gluon propagators,'' {\em Sov. J. Nucl. Phys.} {\bf 45} (1987)
127.

\bibitem{Korchemsky:1986fj}
G.~Korchemsky and A.~Radyushkin, ``Infrared asymptotics of perturbative {QCD}.
  {V}ertex functions,'' {\em Sov. J. Nucl. Phys.} {\bf 45} (1987)
910.

\bibitem{Korchemsky:1987wg}
G.~Korchemsky and A.~Radyushkin, ``{Renormalization of the Wilson Loops Beyond
  the Leading Order},'' {\em Nucl. Phys.} {\bf B283} (1987)
342--364.

\bibitem{Korchemsky:1988hd}
G.~P. Korchemsky, ``Sudakov form-factor in {QCD},'' {\em Phys. Lett.} {\bf
  B220} (1989)
629.

\bibitem{Korchemsky:1988si}
G.~P. Korchemsky, ``{Asymptotics of the Altarelli-Parisi-Lipatov Evolution
  Kernels of Parton Distributions},'' {\em Mod. Phys. Lett.} {\bf A4} (1989)
1257--1276.

\bibitem{Collins:1989bt}
J.~C. Collins, ``{Sudakov form-factors},'' {\em Adv. Ser. Direct. High Energy
  Phys.} {\bf 5} (1989) 573--614,
\href{http://www.arXiv.org/abs/hep-ph/0312336}{{\tt hep-ph/0312336}}.

\bibitem{Korchemsky:1991zp}
G.~Korchemsky and A.~Radyushkin, ``{Infrared factorization, Wilson lines and
  the heavy quark limit},'' {\em Phys. Lett.} {\bf B279} (1992) 359--366,
\href{http://www.arXiv.org/abs/hep-ph/9203222}{{\tt hep-ph/9203222}}.

\bibitem{Kidonakis:1998nf}
N.~Kidonakis, G.~Oderda, and G.~F. Sterman, ``{Evolution of color exchange in
  {QCD} hard scattering},'' {\em Nucl. Phys.} {\bf B531} (1998) 365--402,
\href{http://www.arXiv.org/abs/hep-ph/9803241}{{\tt hep-ph/9803241}}.

\bibitem{Kidonakis:1997gm}
N.~Kidonakis and G.~F. Sterman, ``{Resummation for QCD hard scattering},'' {\em
  Nucl. Phys.} {\bf B505} (1997) 321--348,
\href{http://www.arXiv.org/abs/hep-ph/9705234}{{\tt hep-ph/9705234}}.

\bibitem{Kidonakis:1996aq}
N.~Kidonakis and G.~F. Sterman, ``{Subleading logarithms in QCD hard
  scattering},'' {\em Phys. Lett.} {\bf B387} (1996)
867--874.

\bibitem{Kidonakis:2010dk}
N.~Kidonakis, ``{Next-to-next-to-leading soft-gluon corrections for the top
  quark cross section and transverse momentum distribution},'' {\em Phys. Rev.}
  {\bf D82} (2010) 114030,
\href{http://www.arXiv.org/abs/1009.4935}{{\tt 1009.4935}}.

\bibitem{Drummond:2007cf}
J.~Drummond, J.~Henn, G.~Korchemsky, and E.~Sokatchev, ``{On planar gluon
  amplitudes/Wilson loops duality},'' {\em Nucl. Phys.} {\bf B795} (2008)
  52--68,
\href{http://www.arXiv.org/abs/0709.2368}{{\tt 0709.2368}}.

\bibitem{Basso:2007wd}
B.~Basso, G.~P. Korchemsky, and J.~Kotanski, ``{Cusp anomalous dimension in
  maximally supersymmetric Yang- Mills theory at strong coupling},'' {\em Phys.
  Rev. Lett.} {\bf 100} (2008) 091601,
\href{http://www.arXiv.org/abs/0708.3933}{{\tt 0708.3933}}.

\bibitem{Alday:2007hr}
L.~F. Alday and J.~M. Maldacena, ``{Gluon scattering amplitudes at strong
  coupling},'' {\em JHEP} {\bf 0706} (2007) 064,
\href{http://www.arXiv.org/abs/0705.0303}{{\tt 0705.0303}}.

\bibitem{Pestun:2007rz}
V.~Pestun, ``{Localization of gauge theory on a four-sphere and supersymmetric
  Wilson loops},'' {\em Commun. Math. Phys.} {\bf 313} (2012) 71--129,
\href{http://www.arXiv.org/abs/0712.2824}{{\tt 0712.2824}}.

\bibitem{Drukker:2012de}
N.~Drukker, ``{Integrable Wilson loops},''
\href{http://www.arXiv.org/abs/1203.1617}{{\tt 1203.1617}}.

\bibitem{Chien:2011wz}
Y.-T. Chien, M.~D. Schwartz, D.~Simmons-Duffin, and I.~W. Stewart, ``{Jet
  Physics from Static Charges in AdS},'' {\em Phys. Rev.} {\bf D85} (2012)
  045010,
\href{http://www.arXiv.org/abs/1109.6010}{{\tt 1109.6010}}.

\bibitem{Cherednikov:2012qq}
I.~Cherednikov, T.~Mertens, and F.~Van~der Veken, ``{Cusped light-like Wilson
  loops in gauge theories},'' {\em Phys. Part. Nucl.} {\bf 44} (2013) 250--259,
\href{http://www.arXiv.org/abs/1210.1767}{{\tt 1210.1767}}.

\bibitem{Cherednikov:2012yd}
I.~Cherednikov, T.~Mertens, and F.~Van~der Veken, ``{Evolution of cusped
  light-like Wilson loops and geometry of the loop space},'' {\em Phys. Rev.}
  {\bf D86} (2012) 085035,
\href{http://www.arXiv.org/abs/1208.1631}{{\tt 1208.1631}}.

\bibitem{Henn:2013wfa}
J.~M. Henn and T.~Huber, ``{The four-loop cusp anomalous dimension in N=4 super
  Yang-Mills and analytic integration techniques for Wilson line integrals},''
\href{http://www.arXiv.org/abs/1304.6418}{{\tt 1304.6418}}.

\bibitem{Naculich:2011ry}
S.~G. Naculich and H.~J. Schnitzer, ``{Eikonal methods applied to gravitational
  scattering amplitudes},'' {\em JHEP} {\bf 1105} (2011) 087,
\href{http://www.arXiv.org/abs/1101.1524}{{\tt 1101.1524}}.

\bibitem{White:2011yy}
C.~D. White, ``{Factorization Properties of Soft Graviton Amplitudes},'' {\em
  JHEP} {\bf 1105} (2011) 060, \href{http://www.arXiv.org/abs/1103.2981}{{\tt
  1103.2981}}.

\bibitem{Akhoury:2011kq}
R.~Akhoury, R.~Saotome, and G.~Sterman, ``{Collinear and Soft Divergences in
  Perturbative Quantum Gravity},'' {\em Phys. Rev.} {\bf D84} (2011) 104040,
\href{http://www.arXiv.org/abs/1109.0270}{{\tt 1109.0270}}.

\bibitem{Miller:2012an}
D.~Miller and C.~White, ``{The Gravitational cusp anomalous dimension from AdS
  space},'' {\em Phys. Rev.} {\bf D85} (2012) 104034,
\href{http://www.arXiv.org/abs/1201.2358}{{\tt 1201.2358}}.

\bibitem{Beneke:2012xa}
M.~Beneke and G.~Kirilin, ``{Soft-collinear gravity},'' {\em JHEP} {\bf 1209}
  (2012) 066,
\href{http://www.arXiv.org/abs/1207.4926}{{\tt 1207.4926}}.

\bibitem{Collins:1989gx}
J.~C. Collins, D.~E. Soper, and G.~F. Sterman, ``{Factorization of Hard
  Processes in QCD},'' {\em Adv. Ser. Direct. High Energy Phys.} {\bf 5} (1988)
  1--91,
\href{http://www.arXiv.org/abs/hep-ph/0409313}{{\tt hep-ph/0409313}}.

\bibitem{Korchemsky:1992xv}
G.~P. Korchemsky and G.~Marchesini, ``{Structure function for large x and
  renormalization of Wilson loop},'' {\em Nucl. Phys.} {\bf B406} (1993)
  225--258,
\href{http://www.arXiv.org/abs/hep-ph/9210281}{{\tt hep-ph/9210281}}.

\bibitem{Korchemsky:1993uz}
G.~P. Korchemsky and G.~Marchesini, ``{Resummation of large infrared
  corrections using Wilson loops},'' {\em Phys. Lett.} {\bf B313} (1993)
433--440.

\bibitem{Catani:1996yz}
S.~Catani, M.~L. Mangano, P.~Nason, and L.~Trentadue, ``{The Resummation of
  soft gluons in hadronic collisions},'' {\em Nucl. Phys.} {\bf B478} (1996)
  273--310,
\href{http://www.arXiv.org/abs/hep-ph/9604351}{{\tt hep-ph/9604351}}.

\bibitem{Oderda:1999kr}
G.~Oderda, ``{Dijet rapidity gaps in photoproduction from perturbative QCD},''
  {\em Phys. Rev.} {\bf D61} (2000) 014004,
\href{http://www.arXiv.org/abs/hep-ph/9903240}{{\tt hep-ph/9903240}}.

\bibitem{Bonciani:1998vc}
R.~Bonciani, S.~Catani, M.~L. Mangano, and P.~Nason, ``{NLL resummation of the
  heavy quark hadroproduction cross-section},'' {\em Nucl. Phys.} {\bf B529}
  (1998) 424--450,
\href{http://www.arXiv.org/abs/hep-ph/9801375}{{\tt hep-ph/9801375}}.

\bibitem{Beneke:2009rj}
M.~Beneke, P.~Falgari, and C.~Schwinn, ``{Soft radiation in heavy-particle pair
  production: all- order colour structure and two-loop anomalous dimension},''
  {\em Nucl. Phys.} {\bf B828} (2010) 69--101,
\href{http://www.arXiv.org/abs/0907.1443}{{\tt 0907.1443}}.

\bibitem{Beneke:2009ye}
M.~Beneke, M.~Czakon, P.~Falgari, A.~Mitov, and C.~Schwinn, ``{Threshold
  expansion of the $gg(q\bar{q}) \to Q\bar{Q} + X$ cross section at ${\cal
  O}(\alpha_s^4)$.},'' {\em Phys. Lett.} {\bf B690} (2010) 483--490,
\href{http://www.arXiv.org/abs/0911.5166}{{\tt 0911.5166}}.

\bibitem{Ahrens:2010zv}
V.~Ahrens, A.~Ferroglia, M.~Neubert, B.~D. Pecjak, and L.~L. Yang,
  ``{Renormalization-Group Improved Predictions for Top-Quark Pair Production
  at Hadron Colliders},'' {\em JHEP} {\bf 1009} (2010) 097,
\href{http://www.arXiv.org/abs/1003.5827}{{\tt 1003.5827}}.

\bibitem{Czakon:2013goa}
M.~Czakon, P.~Fiedler, and A.~Mitov, ``{The total top quark pair production
  cross-section at hadron colliders through ${\cal O}(alpha_S^4)$},'' {\em
  Phys. Rev. Lett.} {\bf 110} (2013) 252004,
\href{http://www.arXiv.org/abs/1303.6254}{{\tt 1303.6254}}.

\bibitem{Bauer:2000ew}
C.~W. Bauer, S.~Fleming, and M.~E. Luke, ``{Summing Sudakov logarithms in
  $B\rightarrow X(s \gamma)$ in effective field theory},'' {\em Phys. Rev.}
  {\bf D63} (2000) 014006, \href{http://www.arXiv.org/abs/hep-ph/0005275}{{\tt
  hep-ph/0005275}}.

\bibitem{Bauer:2000yr}
C.~W. Bauer, S.~Fleming, D.~Pirjol, and I.~W. Stewart, ``{An Effective field
  theory for collinear and soft gluons: Heavy to light decays},'' {\em Phys.
  Rev.} {\bf D63} (2001) 114020,
  \href{http://www.arXiv.org/abs/hep-ph/0011336}{{\tt hep-ph/0011336}}.

\bibitem{Bauer:2001ct}
C.~W. Bauer and I.~W. Stewart, ``{Invariant operators in collinear effective
  theory},'' {\em Phys. Lett.} {\bf B516} (2001) 134--142,
  \href{http://www.arXiv.org/abs/hep-ph/0107001}{{\tt hep-ph/0107001}}.

\bibitem{Bauer:2001yt}
C.~W. Bauer, D.~Pirjol, and I.~W. Stewart, ``{Soft collinear factorization in
  effective field theory},'' {\em Phys. Rev.} {\bf D65} (2002) 054022,
  \href{http://www.arXiv.org/abs/hep-ph/0109045}{{\tt hep-ph/0109045}}.

\bibitem{Bauer:2002nz}
C.~W. Bauer, S.~Fleming, D.~Pirjol, I.~Z. Rothstein, and I.~W. Stewart, ``{Hard
  scattering factorization from effective field theory},'' {\em Phys. Rev.}
  {\bf D66} (2002) 014017, \href{http://www.arXiv.org/abs/hep-ph/0202088}{{\tt
  hep-ph/0202088}}.

\bibitem{Becher:2006nr}
T.~Becher and M.~Neubert, ``{Threshold resummation in momentum space from
  effective field theory},'' {\em Phys. Rev. Lett.} {\bf 97} (2006) 082001,
\href{http://www.arXiv.org/abs/hep-ph/0605050}{{\tt hep-ph/0605050}}.

\bibitem{Becher:2006mr}
T.~Becher, M.~Neubert, and B.~D. Pecjak, ``{Factorization and momentum-space
  resummation in deep- inelastic scattering},'' {\em JHEP} {\bf 01} (2007) 076,
\href{http://www.arXiv.org/abs/hep-ph/0607228}{{\tt hep-ph/0607228}}.

\bibitem{Becher:2007ty}
T.~Becher, M.~Neubert, and G.~Xu, ``{Dynamical Threshold Enhancement and
  Resummation in Drell- Yan Production},'' {\em JHEP} {\bf 07} (2008) 030,
\href{http://www.arXiv.org/abs/0710.0680}{{\tt 0710.0680}}.

\bibitem{Yennie:1961ad}
D.~R. Yennie, S.~C. Frautschi, and H.~Suura, ``{The infrared divergence
  phenomena and high-energy processes},'' {\em Ann. Phys.} {\bf 13} (1961)
379--452.

\bibitem{Sterman:1986aj}
G.~F. Sterman, ``{Summation of Large Corrections to Short Distance Hadronic
  Cross-Sections},'' {\em Nucl. Phys.} {\bf B281} (1987)
310.

\bibitem{Catani:1989ne}
S.~Catani and L.~Trentadue, ``{Resummation of the QCD Perturbative Series for
  Hard Processes},'' {\em Nucl. Phys.} {\bf B327} (1989)
323.

\bibitem{Laenen:2008gt}
E.~Laenen, G.~Stavenga, and C.~D. White, ``{Path integral approach to eikonal
  and next-to-eikonal exponentiation},'' {\em JHEP} {\bf 03} (2009) 054,
\href{http://www.arXiv.org/abs/0811.2067}{{\tt 0811.2067}}.

\bibitem{Sotiropoulos:1993rd}
M.~G. Sotiropoulos and G.~F. Sterman, ``{Color exchange in near forward hard
  elastic scattering},'' {\em Nucl. Phys.} {\bf B419} (1994) 59--76,
\href{http://www.arXiv.org/abs/hep-ph/9310279}{{\tt hep-ph/9310279}}.

\bibitem{Korchemsky:1993hr}
G.~P. Korchemsky, ``{On Near forward high-energy scattering in QCD},'' {\em
  Phys. Lett.} {\bf B325} (1994) 459--466,
\href{http://www.arXiv.org/abs/hep-ph/9311294}{{\tt hep-ph/9311294}}.

\bibitem{Korchemskaya:1994qp}
I.~Korchemskaya and G.~Korchemsky, ``{High-energy scattering in QCD and cross
  singularities of Wilson loops},'' {\em Nucl. Phys.} {\bf B437} (1995)
  127--162,
\href{http://www.arXiv.org/abs/hep-ph/9409446}{{\tt hep-ph/9409446}}.

\bibitem{Korchemskaya:1996je}
I.~Korchemskaya and G.~Korchemsky, ``{Evolution equation for gluon Regge
  trajectory},'' {\em Phys. Lett.} {\bf B387} (1996) 346--354,
\href{http://www.arXiv.org/abs/hep-ph/9607229}{{\tt hep-ph/9607229}}.

\bibitem{Balitsky:1995ub}
I.~Balitsky, ``{Operator expansion for high-energy scattering},'' {\em Nucl.
  Phys.} {\bf B463} (1996) 99--160,
\href{http://www.arXiv.org/abs/hep-ph/9509348}{{\tt hep-ph/9509348}}.

\bibitem{Kovchegov:1996ty}
Y.~V. Kovchegov, ``{NonAbelian Weizsacker-Williams field and a two-dimensional
  effective color charge density for a very large nucleus},'' {\em Phys. Rev.}
  {\bf D54} (1996) 5463--5469,
\href{http://www.arXiv.org/abs/hep-ph/9605446}{{\tt hep-ph/9605446}}.

\bibitem{Balitsky:2001gj}
I.~Balitsky, ``{High-energy QCD and Wilson lines},''
\href{http://www.arXiv.org/abs/hep-ph/0101042}{{\tt hep-ph/0101042}}.

\bibitem{Balitsky:2009yp}
I.~Balitsky and G.~A. Chirilli, ``{High-energy amplitudes in N=4 SYM in the
  next-to-leading order},'' {\em Phys. Lett.} {\bf B687} (2010) 204--213,
\href{http://www.arXiv.org/abs/0911.5192}{{\tt 0911.5192}}.

\bibitem{JalilianMarian:1996xn}
J.~Jalilian-Marian, A.~Kovner, L.~D. McLerran, and H.~Weigert, ``{The Intrinsic
  glue distribution at very small x},'' {\em Phys. Rev.} {\bf D55} (1997)
  5414--5428,
\href{http://www.arXiv.org/abs/hep-ph/9606337}{{\tt hep-ph/9606337}}.

\bibitem{Gardi:2006rp}
E.~Gardi, J.~Kuokkanen, K.~Rummukainen, and H.~Weigert, ``{Running coupling and
  power corrections in nonlinear evolution at the high-energy limit},'' {\em
  Nucl. Phys.} {\bf A784} (2007) 282--340,
\href{http://www.arXiv.org/abs/hep-ph/0609087}{{\tt hep-ph/0609087}}.

\bibitem{DelDuca:2011xm}
V.~Del~Duca, C.~Duhr, E.~Gardi, L.~Magnea, and C.~D. White, ``{An infrared
  approach to Reggeization},'' {\em Phys. Rev.} {\bf D85} (2012) 071104,
\href{http://www.arXiv.org/abs/1108.5947}{{\tt 1108.5947}}.

\bibitem{DelDuca:2011ae}
V.~Del~Duca, C.~Duhr, E.~Gardi, L.~Magnea, and C.~D. White, ``{The Infrared
  structure of gauge theory amplitudes in the high-energy limit},'' {\em JHEP}
  {\bf 1112} (2011) 021,
\href{http://www.arXiv.org/abs/1109.3581}{{\tt 1109.3581}}.

\bibitem{Kovchegov:2012mbw}
Y.~V. Kovchegov and E.~Levin,
``{Quantum chromodynamics at high energy},''.

\bibitem{Mueller:1993rr}
A.~H. Mueller, ``{Soft gluons in the infinite momentum wave function and the
  BFKL pomeron},'' {\em Nucl. Phys.} {\bf B415} (1994)
373--385.

\bibitem{Melville:2013qca}
S.~Melville, S.~Naculich, H.~Schnitzer, and C.~White, ``{Wilson line approach
  to gravity in the high energy limit},''
\href{http://www.arXiv.org/abs/1306.6019}{{\tt 1306.6019}}.

\bibitem{Akhoury:2013yua}
R.~Akhoury, R.~Saotome, and G.~Sterman, ``{High Energy Scattering in
  Perturbative Quantum Gravity at Next to Leading Power},''
\href{http://www.arXiv.org/abs/1308.5204}{{\tt 1308.5204}}.

\bibitem{Ware:2013zja}
J.~Ware, R.~Saotome, and R.~Akhoury, ``{Construction of an asymptotic S matrix
  for perturbative quantum gravity},''
\href{http://www.arXiv.org/abs/1308.6285}{{\tt 1308.6285}}.

\bibitem{Gatheral:1983cz}
J.~G.~M. Gatheral, ``{Exponentiation of eikonal cross-sections in nonabelian
  gauge theories},'' {\em Phys. Lett.} {\bf B133} (1983)
90.

\bibitem{Frenkel:1984pz}
J.~Frenkel and J.~C. Taylor, ``{Nonabelian eikonal exponentiation},'' {\em
  Nucl. Phys.} {\bf B246} (1984)
231.

\bibitem{Sterman:1981jc}
G.~F. Sterman, ``Infrared divergences in perturbative {QCD}. (talk),'' {\em AIP
  Conf. Proc.}
22--40.

\bibitem{Aybat:2006wq}
S.~M. Aybat, L.~J. Dixon, and G.~F. Sterman, ``{The two-loop anomalous
  dimension matrix for soft gluon exchange},'' {\em Phys. Rev. Lett.} {\bf 97}
  (2006) 072001,
\href{http://www.arXiv.org/abs/hep-ph/0606254}{{\tt hep-ph/0606254}}.

\bibitem{Aybat:2006mz}
S.~M. Aybat, L.~J. Dixon, and G.~F. Sterman, ``{The two-loop soft anomalous
  dimension matrix and resummation at next-to-next-to leading pole},'' {\em
  Phys. Rev.} {\bf D74} (2006) 074004,
\href{http://www.arXiv.org/abs/hep-ph/0607309}{{\tt hep-ph/0607309}}.

\bibitem{Ferroglia:2009ep}
A.~Ferroglia, M.~Neubert, B.~D. Pecjak, and L.~L. Yang, ``{Two-loop divergences
  of scattering amplitudes with massive partons},'' {\em Phys. Rev. Lett.} {\bf
  103} (2009) 201601,
\href{http://www.arXiv.org/abs/0907.4791}{{\tt 0907.4791}}.

\bibitem{Ferroglia:2009ii}
A.~Ferroglia, M.~Neubert, B.~D. Pecjak, and L.~L. Yang, ``{Two-loop divergences
  of massive scattering amplitudes in non-abelian gauge theories},'' {\em JHEP}
  {\bf 11} (2009) 062,
\href{http://www.arXiv.org/abs/0908.3676}{{\tt 0908.3676}}.

\bibitem{Mitov:2010xw}
A.~Mitov, G.~F. Sterman, and I.~Sung, ``{Computation of the Soft Anomalous
  Dimension Matrix in Coordinate Space},'' {\em Phys. Rev.} {\bf D82} (2010)
  034020,
\href{http://www.arXiv.org/abs/1005.4646}{{\tt 1005.4646}}.

\bibitem{Becher:2009cu}
T.~Becher and M.~Neubert, ``{Infrared singularities of scattering amplitudes in
  perturbative QCD},'' {\em Phys. Rev. Lett.} {\bf 102} (2009) 162001,
\href{http://www.arXiv.org/abs/0901.0722}{{\tt 0901.0722}}.

\bibitem{Becher:2009qa}
T.~Becher and M.~Neubert, ``{On the Structure of Infrared Singularities of
  Gauge-Theory Amplitudes},'' {\em JHEP} {\bf 06} (2009) 081,
\href{http://www.arXiv.org/abs/0903.1126}{{\tt 0903.1126}}.

\bibitem{Gardi:2009qi}
E.~Gardi and L.~Magnea, ``{Factorization constraints for soft anomalous
  dimensions in QCD scattering amplitudes},'' {\em JHEP} {\bf 0903} (2009) 079,
  \href{http://www.arXiv.org/abs/0901.1091}{{\tt 0901.1091}}.

\bibitem{Dixon:2009ur}
L.~J. Dixon, E.~Gardi, and L.~Magnea, ``{On soft singularities at three loops
  and beyond},'' {\em JHEP} {\bf 02} (2010) 081,
\href{http://www.arXiv.org/abs/0910.3653}{{\tt 0910.3653}}.

\bibitem{Ahrens:2012qz}
V.~Ahrens, M.~Neubert, and L.~Vernazza, ``{Structure of Infrared Singularities
  of Gauge-Theory Amplitudes at Three and Four Loops},'' {\em JHEP} {\bf 1209}
  (2012) 138,
\href{http://www.arXiv.org/abs/1208.4847}{{\tt 1208.4847}}.

\bibitem{Naculich:2013xa}
S.~G. Naculich, H.~Nastase, and H.~J. Schnitzer, ``{All-loop infrared-divergent
  behavior of most-subleading-color gauge-theory amplitudes},'' {\em JHEP} {\bf
  1304} (2013) 114,
\href{http://www.arXiv.org/abs/1301.2234}{{\tt 1301.2234}}.

\bibitem{Caron-Huot:2013fea}
S.~Caron-Huot, ``{When does the gluon reggeize?},''
\href{http://www.arXiv.org/abs/1309.6521}{{\tt 1309.6521}}.

\bibitem{Gardi:2010rn}
E.~Gardi, E.~Laenen, G.~Stavenga, and C.~D. White, ``{Webs in multiparton
  scattering using the replica trick},'' {\em JHEP} {\bf 1011} (2010) 155,
  \href{http://www.arXiv.org/abs/1008.0098}{{\tt 1008.0098}}.

\bibitem{Gardi:2011wa}
E.~Gardi and C.~D. White, ``{General properties of multiparton webs: Proofs
  from combinatorics},'' {\em JHEP} {\bf 1103} (2011) 079,
  \href{http://www.arXiv.org/abs/1102.0756}{{\tt 1102.0756}}.

\bibitem{Gardi:2011yz}
E.~Gardi, J.~M. Smillie, and C.~D. White, ``{On the renormalization of
  multiparton webs},'' {\em JHEP} {\bf 1109} (2011) 114,
\href{http://www.arXiv.org/abs/1108.1357}{{\tt 1108.1357}}.

\bibitem{Gardi:2013ita}
E.~Gardi, J.~M. Smillie, and C.~D. White, ``{The Non-Abelian Exponentiation
  theorem for multiple Wilson lines},''
\href{http://www.arXiv.org/abs/1304.7040}{{\tt 1304.7040}}.

\bibitem{Mitov:2010rp}
A.~Mitov, G.~Sterman, and I.~Sung, ``{Diagrammatic Exponentiation for Products
  of Wilson Lines},'' {\em Phys. Rev.} {\bf D82} (2010) 096010,
  \href{http://www.arXiv.org/abs/1008.0099}{{\tt 1008.0099}}.

\bibitem{integrals}
E.~Gardi, ``{From webs to polylogarithms},'' {\em In preparation}.

\bibitem{integrals2}
E.~Gardi, G.~Falconi, M.~Harley, L.~Magnea, and C.~D. White, ``{Calculating
  three-loop webs},'' {\em In preparation}.

\bibitem{Dukes:2013wa}
M.~Dukes, E.~Gardi, E.~Steingrimsson, and C.~D. White, ``{Web worlds,
  web-colouring matrices, and web-mixing matrices},'' {\em J. Comb. Theory Ser.
  A} {\bf 120} (2013) 1012--1037,
\href{http://www.arXiv.org/abs/1301.6576}{{\tt 1301.6576}}.

\bibitem{DelDuca:1999rs}
V.~Del~Duca, L.~J. Dixon, and F.~Maltoni, ``{New color decompositions for gauge
  amplitudes at tree and loop level},'' {\em Nucl. Phys.} {\bf B571} (2000)
  51--70,
\href{http://www.arXiv.org/abs/hep-ph/9910563}{{\tt hep-ph/9910563}}.

\bibitem{Mazur}
D.~R. Mazur, ``{Combinatorics: A Guided Tour},'' {\em Mathematical Association
  of America} (2010).

\bibitem{Brightwell}
G.~Brightwell and P.~Winkler, ``{Counting linear extensions},'' {\em Order}
  {\bf 8} (1991)
225--242.

\bibitem{Sury}
B.~Sury, T.~Wang, and F.~Z. Zhao, ``Identities involving reciprocals of
  binomial coefficients,'' {\em Journal of Integer Sequences} {\bf 7} (2004)
  04.2.8.

\end{thebibliography}\endgroup
\end{document}